\documentclass[journal]{IEEEtran}

\usepackage{etex}

\ifCLASSINFOpdf
\else
\fi

\usepackage{url}
\usepackage{amsmath, amssymb, bm, cite, epsfig}
\usepackage{epstopdf}
\usepackage{graphicx}
\usepackage{array}
\usepackage{multirow}
\renewcommand{\arraystretch}{1.5}
\usepackage{amsfonts}
\usepackage{tabularx} 
\usepackage{tabu}
\usepackage{pbox}
\usepackage{makecell}

\usepackage{booktabs}
\usepackage{ctable}
\usepackage{xcolor,colortbl}
\definecolor{Gray}{gray}{0.85}
\definecolor{LightCyan}{rgb}{0.8,1,1}
\usepackage[margin=0.568in]{geometry}

\def\beq{\begin{equation}}
\def\eeq{\end{equation}}
\def\beqa{\begin{eqnarray}}
\def\eeqa{\end{eqnarray}}
\def\beqan{\begin{eqnarray*}}
\def\eeqan{\end{eqnarray*}}

\def\diag{\mathop{\mathrm{diag}}}

\def\PL{\mathrm{PL}}
\def\dB{\mathrm{dB}}

\def\tm1{t\! - \! 1}
\def\tp1{t\! + \! 1}

\def\PL{\mathrm{PL}}
\def\dB{\mathrm{dB}}
\def\PLE{\mathrm{PLE}}
\def\FSPL{\mathrm{FSPL}}
\def\log{\mathrm{log}}
\def\ABG{\mathrm{ABG}}
\def\CI{\mathrm{CI}}
\def\CIF{\mathrm{CIF}}
\def\m{\mathrm{m}}

\usepackage{tikz}
\usetikzlibrary{calc}

\begin{document}
\pagenumbering{gobble}
\title{Investigation of Prediction Accuracy, Sensitivity, and Parameter Stability of Large-Scale Propagation Path Loss Models for 5G Wireless Communications}

\author{\IEEEauthorblockN{Shu Sun, \textit{Student Member, IEEE}, Theodore S. Rappaport, \textit{Fellow, IEEE}, Timothy A. Thomas, \textit{Member, IEEE}, Amitava Ghosh, \textit{Fellow, IEEE}, Huan C.  Nguyen, \textit{Member, IEEE}, Istv$\acute{a}$n Z. Kov$\acute{a}$cs, \textit{Member, IEEE}, Ignacio Rodriguez, \textit{Student Member, IEEE}, Ozge Koymen, \textit{Member, IEEE}, and Andrzej Partyka, \textit{Member, IEEE}}

\thanks{This material is based upon work supported by the NYU WIRELESS Industrial Affiliates: AT\&T, CableLabs, Ericsson, Huawei, Intel Corporation, InterDigital Inc., Keysight Technologies, L3 Communications, Nokia, National Instruments, Qualcomm Technologies, Samsung Corporation, SiBeam, Straight Path Communications, Cablevision, UMC, and XO Communications. This work is also supported by the GAANN Fellowship Program and three National Science Foundation Grants (NSF Accelerating Innovative Research EAGER (Award Number:1555332), NeTS Medium (Award Number:1302336), and NeTS Small (Award Number:1320472)). The authors would like to thank Mathew K. Samimi and George R. MacCartney, Jr. at NYU WIRELESS for their contribution to this work. Initial results related to this work have been published in\cite{Mac15:Access,Tho16:VTC,Sun16:VTC,Sun16:EuCAP}.

S. Sun and T. S. Rappaport are with NYU WIRELESS and Tandon School of Engineering, New York University, Brooklyn, NY 11201, USA (e-mail: ss7152@nyu.edu; tsr@nyu.edu).

T. A. Thomas and A. Ghosh are with Nokia, Arlington Heights, IL 60004, USA (e-mail: timothy.thomas@nokia.com; amitava.ghosh@nokia.com).

H. Nguyen and I. Rodriguez are with Aalborg University, Aalborg 9220, Denmark (e-mail: huan.nguyen.ext@nokia.com; irl@es.aau.dk).

I. Z. Kov$\acute{a}$cs is with Nokia, Aalborg 9220, Denmark (e-mail: istvan.kovacs@nokia.com).

O. Koymen and A. Partyka are with Qualcomm R\&D, Bridgewater, NJ 08807, USA (e-mail: okoymen@qti.qualcomm.com; apartyka@qti.qualcomm.com).}
}

\maketitle
\begin{tikzpicture}[remember picture, overlay]
\node at ($(current page.north) + (0,-0.25in)$) {S. Sun \textit{et al.}, \rq\rq{}Investigation of Prediction Accuracy, Sensitivity, and Parameter Stability of Large-Scale Propagation Path Loss};
\node at ($(current page.north) + (0,-0.4in)$) {Models for 5G Wireless Communications,\rq\rq{}~\textit{IEEE Transactions on Vehicular Technology}, vol. 65, no. 5, pp. 2843-2860, May 2016.};
\node at ($(current page.north) + (0,-0.55in)$) {Available at: http://ieeexplore.ieee.org/xpl/articleDetails.jsp?arnumber=7434656.};
\node at ($(current page.north) + (0,-0.70in)$) {};
\end{tikzpicture}
\begin{abstract}
This paper compares three candidate large-scale propagation path loss models for use over the entire microwave and millimeter-wave (mmWave) radio spectrum: the alpha-beta-gamma (ABG) model, the close-in (CI) free-space reference distance model, and the CI model with a frequency-weighted path loss exponent (CIF). Each of these models has been recently studied for use in standards bodies such as 3rd Generation Partnership Project (3GPP) and for use in the design of fifth-generation wireless systems in urban macrocell, urban microcell, and indoor office and shopping mall scenarios. Here, we compare the accuracy and sensitivity of these models using measured data from 30 propagation measurement data sets from 2 to 73 GHz over distances ranging from 4 to 1238 m. A series of sensitivity analyses of the three models shows that the physically based two-parameter CI model and three-parameter CIF model offer computational simplicity, have very similar goodness of fit (i.e., the shadow fading standard deviation), exhibit more stable model parameter behavior across frequencies and distances, and yield smaller prediction error in sensitivity tests across distances and frequencies, when compared to the four-parameter ABG model. Results show the
CI model with a 1-m reference distance is suitable for outdoor environments, while the CIF model is more appropriate for indoor modeling. The CI and CIF models are easily implemented in
existing 3GPP models by making a very subtle modification --- by replacing a floating non-physically based constant with a frequency-dependent constant that represents free-space path loss in the first meter of propagation. This paper shows this subtle change does not change the mathematical form of existing ITU/3GPP models and offers much easier analysis, intuitive appeal, better model parameter stability, and better accuracy in sensitivity tests over a vast range of microwave and mmWave frequencies, scenarios, and distances, while using a simpler model with fewer parameters.
\end{abstract}
\begin{IEEEkeywords}
Millimeter wave, path loss models, prediction accuracy, 5G.
\end{IEEEkeywords}

\IEEEpeerreviewmaketitle
\section{Introduction}
The rapidly increasing demands for higher mobile data rates and ubiquitous data access have led to a spectrum crunch over the traditional wireless communication frequency bands, i.e., below 6 GHz. Innovative technologies such as multiple-input multiple-output\cite{Boc14,And14,Sun14}, and new spectrum allocations in the millimeter-wave (mmWave) frequency bands\cite{Rap13:Access}, are useful to alleviate the current spectrum shortage\cite{5GSIG:WP}, and are driving the development of the fifth-generation (5G) wireless communications. It is necessary to have a good knowledge of the propagation channel characteristics across all microwave and mmWave frequencies in order to conduct accurate and reliable 5G system design. 

Emerging 5G communication systems are expected to employ revolutionary technologies\cite{Ran14}, potential new spectra\cite{Rap15}, and novel architectural concepts\cite{And14,Boc14}, hence it is critical to develop reliable channel models to assist engineers in the design. Channel characterization at both mmWave and centimeter-wave (cmWave) bands has been conducted by many prior researchers\cite{Han:VTC,Han:ICCW}. For instance, wideband non-line-of-sight (NLOS) channels at 9.6, 28.8, and 57.6 GHz in downtown Denver were measured in\cite{Vio88}; Lovnes \textit{et al.} and Smulders \textit{et al.} performed outdoor propagation measurements and modeling at the 60 GHz band in a variety of city streets\cite{Lov94, Smu97}. Over the past few years, a number of measurement campaigns, prototypes, or modeling work for mmWave channels for future mobile communications have been conducted by Nokia\cite{Cudak14,Ghosh14} and Samsung\cite{Hur14,Roh14}; Kyro \textit{et al.} from Aalto University performed channel measurements at 81 GHz to 86 GHz of the E-band for point-to-point communications in a street canyon scenario in Helsinki, Finland\cite{Kyro12}. Additionally, extensive propagation measurements and channel modeling were carried out at 28 GHz, 38 GHz, 60 GHz, and 73 GHz in urban microcell (UMi), urban macrocell (UMa), and indoor hotspot (InH) scenarios\cite{Rap13:Access,RapGut13,Mac14,Rap15:TCOM}. Note that for the UMi scenario, the base station (BS) antenna is at rooftop height, typically 10 m or so above ground as defined in\cite{3GPP:25996,3GPP_LTE}; while for the UMa scenario,  the BS antenna is above rooftop height, typically 25 m or so above ground as defined in\cite{3GPP:25996,3GPP_LTE}. Raw data representing corresponding measured path loss data for the indoor and outdoor measurements were provided in\cite{Mac15:Access} and\cite{Mac15:PL}. Large-scale path loss models at 38 GHz and 60 GHz were published for urban outdoor environments in Austin, Texas\cite{Ben11,Rap_Ben12}. Directional and omnidirectional path loss models in dense urban environments at 28 GHz and 73 GHz were presented in\cite{Mac14:PIMRC,Samimi15:WCL}. Spatial and temporal statistics based on UMi measurements at 28 GHz and 73 GHz were extracted in combination with ray-tracing\cite{Rap15:TCOM,Samimi16:EuCAP}. Two-dimensional (2D) and 3D statistical spatial channel models for across the mmWave bands were developed in\cite{Samimi16:EuCAP,Samimi15:MTT}. 

3rd Generation Partnership Project (3GPP)\cite{3GPP:25996} and WINNER II\cite{WINNER} channel models are the most well-known and widely employed models in industry, containing a diversity of deployment scenarios such as UMi, UMa, indoor office, indoor shopping mall, etc., and they provide key channel parameters including line-of-sight (LOS) probabilities, path loss models, path delays, and path powers. However, the 3GPP and WINNER channel models are only applicable for  frequency bands below 6 GHz; hence all of the modeling methodologies need to be revisited and revised for frequency bands above 6 GHz\cite{Han:ICCW,Han:VTC}. In addition, as discussed subsequently, a three-parameter floating-intercept (\textit{alpha-beta} (AB)) large-scale path loss model was adopted by 3GPP and WINNER, which offers a standard model but lacks solid physical meaning due to its widely varying (floating) modeling parameters when applied in a particular band of frequencies or scenario\cite{Sun16:VTC}. 

This paper investigates three large-scale path loss models that may be used over the microwave and mmWave frequency bands: the alpha-beta-gamma (ABG) model, the close-in (CI) free space reference distance path loss model, and the CI model with a frequency-weighted path loss exponent (CIF)\cite{GRM13:Globecom,Pie12,And95,Sun16:VTC,Mac15:Access,Sun16:EuCAP}, which is a general form of the CI model. The ABG model is shown to be a simple extension of the AB model currently used in 3GPP, where a frequency-dependent floating optimization parameter is added to the AB model. We also show that the CI and CIF models are simpler in form (require fewer parameters), and offer better parameter stability and accuracy through the use of a physically-based close-in reference distance that replaces the floating model parameters of the ABG model. In this work, systematic comparisons between the parameters, shadow fading (SF) standard deviations, and prediction performance of these three models in the UMa, UMi street canyon (SC), InH office, and InH shopping mall (SM) scenarios are provided, using eight sets of measurement data from New York University (NYU), two sets from The University of Texas at Austin (UT Austin), eight sets from Nokia/Aalborg University (AAU), and 12 sets from Qualcomm.

\section{Overview of The Measurement Campaigns}
\subsection{UMa Measurements at Aalborg University}
UMa propagation measurements were performed in Vestby, Aalborg, Denmark, in the 2 GHz, 10 GHz, 18 GHz, and 28 GHz frequency bands in March 2015\cite{Ngu16:VTC,Sun15:GCWS}. Vestby represents a typical medium-sized European city with regular building height and street width, which is approximately 17 m (5 floors) and 20 m, respectively. There were six transmitter (TX) locations, with a TX antenna height of 20 or 25 m. A narrowband continuous wave (CW) signal was transmitted at 10, 18 and 28 GHz, and another CW signal at 2 GHz was always transmitted in parallel and served as a reference. The eight different sets of data at different frequencies were measured at identical locations. The receiver (RX) was mounted on a van, with a height of approximately 2.4 m. It has been shown in\cite{Mac14} that a 4 m RX height may be expected to have comparable path loss to a lower height of 2 m in urban settings, hence the results obtained by using the 2.4 m-high RX in the UMa measurements should be comparable to those using an RX at typical mobile heights. The van was driven at a speed of 20 km/h within the experimental area, and the driving routes were chosen so that they were confined within the HPBW of the TX antennas. The received signal strength and GPS location were recorded at a rate of 20 samples/s using the R\&S TSMW Universal Radio Network Analyzer for the calculation of path loss and TX-RX (T-R) separation distances. The data points were visually classified into LOS and NLOS conditions based on Google Maps.

\subsection{UMa Measurements at UT Austin}
In the summer of 2011, 38 GHz propagation measurements were conducted with four TX locations chosen on buildings at the UT Austin campus\cite{RapGut13,Rap15:TCOM,Sun15:GCWS}, using a spread spectrum sliding correlator channel sounder and directional steerable high-gain horn antennas, with a center carrier frequency of 37.625 GHz, a maximum RF transmit power of 21.2 dBm over an 800 MHz first null-to-null RF bandwidth and a maximum measurable dynamic range of 160 dB. The measurements used narrowbeam TX antennas (7.8$^{\circ}$ azimuth half-power beamwidth (HPBW)) and narrowbeam (7.8$^{\circ}$ azimuth HPBW) or widebeam (49.4$^{\circ}$ azimuth HPBW) RX antennas. Among the four TX sites, three were with heights of 23 m or 36 m, representing the typical heights of base stations in the UMa scenario, and all the RX sites had a height of 1.5 m (representing typical mobile heights). A total of 33 TX-RX location combinations were measured using the narrowbeam RX antenna (with 3D T-R separation distances ranging from 61 m to 930 m) and 15 TX-RX location combinations were measured using the widebeam RX antenna (with 3D T-R separation distances between 70 m and 728 m) for the UMa scenario, where for each TX-RX location combination, power delay profiles (PDPs) for several TX and RX antenna azimuth and elevation pointing angle combinations were recorded. Raw path loss data from the 38 GHz measurement campaign are provided in\cite{Mac15:PL}. This paper uses two measurement data sets comprised of all measurement data using narrowbeam antennas (21 LOS omnidirectional locations, and 12 NLOS locations).

\subsection{UMi and InH Measurements at NYU}
UMi SC and InH office channel measurements were conducted by NYU at 28 GHz and 73 GHz\cite{Rap13:Access,Rap15:TCOM,Mac14,Mac15:Access}, using a 400 Megachips-per-second (Mcps) spread spectrum sliding correlator channel sounder and directional steerable horn antennas at both the TX and RX. The raw data contained in the eight data sets of outdoor and indoor 28 GHz and 73 GHz path loss measurements are given in\cite{Mac15:PL,Mac15:Access}. Detailed information about the measurement equipment, measurement procedures, and some measurement results are available in\cite{Rap13:Access,Rap15:TCOM,Mac14,Mac15:Access}. 

\subsection{UMi and InH Measurements at Qualcomm}
For both the UMi and InH measurements conducted by Qualcomm, a channel sounder operating at 2.9 GHz, 29 GHz, and 61 GHz was used. The time resolution of the channel sounder is approximately 5 ns. Omnidirectional antennas were employed for the 2.9 GHz measurements, while directional antennas with gains of 10 dBi and 20 dBi at 29 GHz and 61 GHz were used for scans in both azimuth 360$^\circ$ and in elevation from -30$^\circ$ to +90$^\circ$. The resultant scan includes 39 slices with a 10 dBi-gain antenna and 331 slices with a 20 dBi-gain antenna. 

For the outdoor measurement campaign by Qualcomm, the environment included office campus around 500 Somerset Corporate Boulevard in Bridgewater, NJ. Two data sets were measured from the site including five multi-level buildings, two parking lots and connecting streets and walkways and nearby large shopping malls, all surrounded by dense rows of trees (a mix of deciduous trees and dense spruce). T-R distances varied between 35 m and 260 m.

The InH office measurements were conducted on two typical office floors, one floor had mostly cubical offices with some closed wall offices centrally located, and the other contained closed wall offices and long corridors. Four data sets were measured at three TX locations were used with about 40 RX locations per TX on average, with a minimum distance of about 5 m, and a maximum distance of 67 m. For the InH SM measurements, three TX locations were used with about 135 RX locations on three floors from which six data sets were obtained, where the minimum distance was about 10 m, and maximum distance was about 275 m. 

\section{Large-Scale Propagation Path Loss Models}
The ABG, CI and CIF path loss models are multi-frequency statistical (i.e., stochastic) models that describe large-scale propagation path loss over distance at all relevant frequencies in a certain scenario\cite{Mac15:Access,Sun16:EuCAP}. It will be noted that the CI and CIF models have a very similar form compared to the existing 3GPP path loss model (i.e., the floating-intercept, or AB model)\cite{3GPP:25996}, where one merely needs to substitute the floating constant (which has been shown to vary substantially across different measurements, frequencies and scenarios\cite{Sun16:VTC,Tho16:VTC}) with a free-space constant that is a function of frequency based on a 1 m standard free space reference distance. As shown subsequently, this subtle change provides a frequency-dependent term while yielding greater prediction accuracy and better parameter stability when using the models outside of the range of the measured data set from which the models are developed.

We note that testing the efficacy of a path loss model outside of the range for which measurements are originally collected and used to solve for model parameters is a critical, but often ignored, test. Testing model accuracy and parameter stability is needed since engineers will inevitably require propagation models for new applications, distances, or scenarios not originally contemplated in the original experiments used to build the path loss model. For future 5G wireless system level and link layer analysis and simulation in new spectrum bands, where new types of directional antennas, umbrella cells, repeater architectures, and new regulations and network topologies are used\cite{Rap15}, it is critical to know that a chosen model can be used in new scenarios while still exhibiting parameter stability, accuracy, and usefulness beyond the limited original number of field measurements. This paper offers such sensitivity and analysis when comparing the three candidate 5G stochastic path loss models.

The equation for the ABG model is given by~\eqref{ABG1}\cite{3GPP_LTE}:
\begin{equation}\label{ABG1}
\begin{split}
\PL^{\ABG}(f,d)[\dB]=&10\alpha \log_{10}\left(\frac{d}{1~m}\right)+\beta\\
&+10\gamma \log_{10}\left(\frac{f}{1~GHz}\right)+\chi_{\sigma}^{\ABG} \text{,}
\\ &\text{ where } d\geq\textrm{ 1 m}
\end{split}
\end{equation}

\noindent where $\PL^{\ABG}(f,d)$ denotes the path loss in dB over frequency and distance, $\alpha$ and $\gamma$ are coefficients showing the dependence of path loss on distance and frequency, respectively, $\beta$ is an optimized offset value for path loss in dB, $d$ is the 3D T-R separation distance in meters, $f$ is the carrier frequency in gigahertz (GHz), and $\chi_{\sigma}^{\ABG}$ is a zero-mean Gaussian random variable with a standard deviation $\sigma$ in dB describing large-scale signal fluctuations (i.e., shadowing) about the mean path loss over distance and frequency. Note that the ABG model has three model parameters for determining mean path loss over distance and frequency, as well as the shadowing standard deviation (a total of four parameters). When used at a single frequency, the ABG model reverts to the existing 3GPP floating-intercept (AB) model with three parameters with $\gamma$ set to 0 or 2\cite{WINNER,Rap15:TCOM,GRM13:Globecom}. The ABG model parameters $\alpha$, $\beta$, $\gamma$, and $\sigma$ are obtained from measured data using the closed-form solutions that minimize the SF standard deviation, as shown in the Appendix. 

The equation for the CI model is given by~\eqref{CI1}\cite{Rappaport:Wireless2nd}:
\begin{equation}\label{CI1}
\begin{split}
\PL^{\CI}(f,d)[\dB]=&\FSPL(f, d_0)[\dB]+10n\log_{10}\left(d/d_0\right)+\chi_{\sigma}^{\CI} \text{,}
\\ &\text{ where } d\geq d_0
\end{split}
\end{equation}

\noindent where $f$ is also in GHz (for both the CI and CIF models), $d_0$ is the close-in free space reference distance, $n$ denotes the path loss exponent (PLE), and $\chi_{\sigma}^{\CI}$ is a zero-mean Gaussian random variable with a standard deviation $\sigma$ in dB. Whereas the ABG model requires four parameters, the CI model only requires one parameter, the PLE, to determine the mean path loss with distance and frequency, and uses a total of two parameters (the PLE $n$ and $\chi_{\sigma}^{\CI}$). A useful property of~\eqref{CI1} is that 10$n$ describes path loss in dB in terms of decades of distances beginning at $d_0$ (making it very easy to compute power over distance in one's mind when $d_0$ is set to 1 m\cite{Rap15:TCOM,Sun16:VTC,Sun16:EuCAP}). In~\eqref{CI1}, $d$ is the 3D T-R separation distance, and $\FSPL(f, d_0)$ denotes the free space path loss in dB at a T-R separation distance of $d_0$ at the carrier frequency $f$:
\begin{equation}\label{FSPL}
\FSPL(f, d_0)[\dB]=20\log_{10}\left(\frac{4\pi fd_0\times 10^9}{c}\right)
\end{equation}

\noindent where $c$ is the speed of light. Note that the CI model inherently has an intrinsic frequency dependency of path loss already embedded within the FSPL term. The PLE model parameter in~\eqref{CI1} is obtained by first removing the FSPL given by~\eqref{FSPL} from the path loss on the left side of~\eqref{CI1} for all measured data points across all frequencies, and then calculating the single PLE jointly for multiple frequencies, as detailed in the Appendix and\cite{Rap15:TCOM}. The CI model in (2) can be written in the 3GPP form\cite{3GPP_LTE} as:
\begin{equation*}\label{CI2}
\begin{split}
\PL^{\CI}(f,d)[\dB]=&\FSPL(f, d_0)[\dB]+10n\log_{10}\left(d/d_0\right)+\chi_{\sigma}^{\CI}
\\ =&10n\log_{10}\left(\frac{d}{d_0}\right)+20\log_{10}\left(\frac{4\pi d_0\times 10^9}{c}\right)
\\&+20\log_{10}\left(f\right)+\chi_{\sigma}^{\CI}
\\=&10n\log_{10}\left(\frac{d}{d_0}\right)+\eta+20\log_{10}\left(f\right)+\chi_{\sigma}^{\CI}\text{,}
\\ &\text{ where } d\geq d_0\text{,}\text{ and } \eta=20\log_{10}\left(\frac{4\pi d_0\times 10^9}{c}\right)
\end{split}
\end{equation*}

The choice of $d_0$ = 1 m as the close-in free space reference distance is shown here to provide excellent parameter stability and model accuracy for outdoor UMi and UMa, and indoor channels across a vast range of microwave and mmWave frequencies, and creates a standardized modeling approach. While the choice of a close-in reference distance of 1 m may be in the near-field of large antenna arrays, the error caused by this in practical wireless system design is negligible, and is more realistic than the ABG model, as shown subsequently and in\cite{Rap15:TCOM}.

A recent path loss model also suitable for multi-frequency modeling follows as a more general form of the CI model, and is called the CIF model, given by Eq.~\eqref{CIF1} when $d_0$ = 1 m\cite{Mac15:Access}:
\begin{equation}\label{CIF1}
\begin{split}
\PL^{\CIF}(f,d)[\dB]=&\FSPL(f, 1~\m)[\dB]+\\
&10n\bigg(1+b\Big(\frac{f-f_0}{f_0}\Big)\bigg)\log_{10}\left(d\right)+\chi_{\sigma}^{\CIF} \text{,}
\\ &\text{ where } d\geq\textrm{ 1 m}
\end{split}
\end{equation}

\noindent where $n$ denotes the distance dependence of path loss (similar to the PLE in the CI model), and $b$ is a model parameter that captures the amount of linear frequency dependence of path loss about the weighted average of all frequencies considered in the model. The CIF model in (4) can also be written in the 3GPP form [26] as:
\begin{equation*}\label{CIF2}
\begin{split}
\PL&^{\CIF}(f,d)[\dB]
\\=&\FSPL(f, 1~\m)[\dB]+10n\bigg(1+b\Big(\frac{f-f_0}{f_0}\Big)\bigg)\log_{10}\left(d\right)+\chi_{\sigma}^{\CIF}
\\=&10n\bigg(1+b\Big(\frac{f-f_0}{f_0}\Big)\bigg)\log_{10}\left(d\right)+20\log_{10}\left(\frac{4\pi\times 10^9}{c}\right)
\\&+20\log_{10}\left(f\right)+\chi_{\sigma}^{\CIF}
\\=&10n\bigg(1+b\Big(\frac{f-f_0}{f_0}\Big)\bigg)\log_{10}\left(d\right)+\eta+20\log_{10}\left(f\right)+\chi_{\sigma}^{\CIF}\text{,}
\\ &\text{ for } d\geq 1\text{ m, }\text{ and } \eta=20\log_{10}\left(\frac{4\pi\times 10^9}{c}\right)=32.4~\dB
\end{split}
\end{equation*}

The parameter $f_0$ is the average frequency calculated by~\eqref{CIF_f_0} that is an input parameter computed from the measurement set used to form the model, and serves as the balancing point for the linear frequency dependence of the PLE:
\begin{equation}\label{CIF_f_0}
f_0 = \frac{\sum_{k=1}^{K}f_k N_k}{\sum_{k=1}^{K}N_k}
\end{equation}
where $K$ is the number of unique frequencies, $N_k$ is the number of path loss data points corresponding to the $k^{th}$ frequency $f_k$, and $\chi_{\sigma}^{\CIF}$ in~\eqref{CIF1} is a zero-mean Gaussian random variable with a standard deviation $\sigma$ in dB that describes large-scale shadowing. Note that the calculated $f_0$ is rounded to the nearest integer in GHz in this work. The CIF model reverts to the CI model for the single frequency case (when $f_0$ is equal to the single frequency $f$) or when $b=0$ (i.e., when there is no frequency dependence on path loss, besides that which occurs in the first meter of free space propagation). As shown subsequently, UMa channels modeled by CIF have a value of $b$ very close to zero, indicating that almost all of the frequency-dependent effects are incorporated in the first meter of free space propagation\cite{Rap15:TCOM,Mac15:Access}. 

The CI and CIF models provide a close-in free space anchor point which assures that the path loss model (regardless of transmit power) always has a physical tie and continuous relationship to the transmitted power over distance, whereas the AB and ABG models use a floating constant based on a fit to the data, without consideration for the close-in free space propagation that always occurs in practice near an antenna out in the open (this implies that particular measured path loss values could greatly impact and skew the ABG path loss model parameters, since there is not a physical anchor to assure that close-in free space transmission occurs in the first meter of propagation from the TX antenna). The CI and CIF models are therefore based on fundamental principles of wireless propagation, dating back to Friis and Bullington, where the PLE parameter offers insight into path loss based on the environment, having a PLE value of 2 in free space (as shown by Friis) and a value of 4 for the asymptotic two-ray ground bounce propagation model (as shown by Bullington)\cite{Rappaport:Wireless2nd}. Previous UHF (Ultra-High Frequency) and microwave models used a close-in reference distance of 1 km or 100 m since BS towers were tall without any nearby obstructions, and inter-site distances were on the order of many kilometers for those frequency bands\cite{Rappaport:Wireless2nd,Hata:TVT80}. We use $d_0$ = 1 m in 5G path loss models since coverage distances will be shorter at higher frequencies. Furthermore, with future small cells, BSs are likely to be mounted closer to obstructions~\cite{Rap13:Access,Rap15:TCOM}. The CI and CIF $d_0$ =1 m reference distance is a suggested standard that ties the true transmitted power or path loss to a convenient close-in distance, as suggested in\cite{Rap15:TCOM}. Standardizing to a reference distance of 1 m makes comparisons of measurements and models simple, and provides a standard definition for the PLE, while enabling intuition and rapid computation of path loss. Now we show with measured data that the 1 m reference is very effective for large-scale path loss modeling across a vast range of frequencies. 

As discussed in\cite{Rap15:TCOM}, emerging mmWave mobile systems will have very few users within a few meters of the BS antenna (in fact, no users are likely to be in the near field, since transmitters will be mounted on a lamppost or ceiling), and users in the near field will have strong signals or will be power-controlled compared to typical users much farther from the transmitter such that any path loss error in the near field (between 1 m and the Fraunhofer distance) will be very minor, and so much smaller than the dynamic range of signals experienced by users in a commercial system.

\begin{table*}
\renewcommand{\arraystretch}{1.4}
\begin{center}
\caption{Parameters in the CI and CI-opt path loss models in UMa and UMi scenarios. Freq. Range denotes frequency range. \# of Data Points represents the number of data points after distance binning and path loss thresholding. Dist. Range denotes distance range, CI-opt represents the CI model with an optimized free space reference distance $d_0$. $\Delta_\sigma$ denotes the difference in the SF standard deviation between the CI and CI-opt models.}~\label{tbl:UMa_CI}
\begin{tabular}{|cc|c|c|c|c|c|c|c|c|c|}
\hline 
 \multicolumn{2}{|c|}{Sce.} & Env.& \makecell{Freq.\\Range\\(GHz)} & \makecell{\# of\\Data \\Points} &\makecell{Dist. \\Range\\(m)}&Model & \makecell{$\PLE$} & $d_0 (m)$ & \makecell{$\sigma$ \\($\dB$)} & \makecell{$\Delta_\sigma$ \\($\dB$)} \\ \Xcline{1-11}{1.5pt}
 \multicolumn{2}{|c|}{\multirow{24}{*}{\makecell{UMa}}} & \multirow{12}{*}{LOS}& \multirow{2}{*}{2} & \multirow{2}{*}{253} & \multirow{2}{*}{60-564} & CI-opt & 2.1 & 6.2 & 1.7 & \multirow{2}{*}{0.0}\\ \cline{7-10}
 & & & & & &  CI & 2.0 & 1 & 1.7 & \\ \cline{4-11}
 & & & \multirow{2}{*}{10}&  \multirow{2}{*}{253} & \multirow{2}{*}{60-564} & CI-opt & 2.0 & 0.1 & 3.1 & \multirow{2}{*}{0.0}\\ \cline{7-10}
& & & & & & CI & 2.0 & 1 & 3.1 & \\ \cline{4-11}
 & & & \multirow{2}{*}{18}&  \multirow{2}{*}{253} & \multirow{2}{*}{60-564} & CI-opt & 2.1 & 14.7 & 2.0 & \multirow{2}{*}{0.0}\\ \cline{7-10} 
 & & & & & & CI & 2.0 & 1 & 2.0 & \\ \cline{4-11}
 & & & \multirow{2}{*}{28}&  \multirow{2}{*}{253} & \multirow{2}{*}{60-564} & CI-opt & 2.0 & 50.0 & 2.3 & \multirow{2}{*}{0.0}\\ \cline{7-10} 
 & & & & & & CI & 2.0 & 1 & 2.3 & \\ \cline{4-11}
& & & \multirow{2}{*}{38}&  \multirow{2}{*}{20} & \multirow{2}{*}{70-930} & CI-opt & 1.7 & 32.9 & 3.4  & \multirow{2}{*}{0.1}\\ \cline{7-10} 
& & & & & & CI & 1.9 & 1 & 3.5 & \\ \cline{4-11}
& & & \multirow{2}{*}{2-38}&  \multirow{2}{*}{1032} & \multirow{2}{*}{60-930} & CI-opt & 2.0 & 0.1 & 2.4 & \multirow{2}{*}{0.0}\\ \cline{7-10} 
& & & & & & CI & 2.0 & 1 & 2.4 & \\ \cline{3-11}
 & & \multirow{12}{*}{NLOS}  & \multirow{2}{*}{2} &  \multirow{2}{*}{583} & \multirow{2}{*}{74-1238} & CI-opt & 3.3 & 10.0 & 3.2 & \multirow{2}{*}{0.3}\\ \cline{7-10} 
& & & & & & CI & 2.8 & 1 & 3.5 & \\ \cline{4-11}
&  & & \multirow{2}{*}{10}&  \multirow{2}{*}{581} & \multirow{2}{*}{74-1238} & CI-opt & 3.4 & 4.3 & 4.0 & \multirow{2}{*}{0.1}\\ \cline{7-10} 
 & & & & & & CI & 3.1 & 1 & 4.1 & \\ \cline{4-11}
& & & \multirow{2}{*}{18}&  \multirow{2}{*}{468} & \multirow{2}{*}{78-1032} & CI-opt & 3.2 & 2.2 & 4.4 & \multirow{2}{*}{0.1}\\ \cline{7-10} 
& & & & & & CI & 3.0 & 1 & 4.5 & \\ \cline{4-11}
& & & \multirow{2}{*}{28}&  \multirow{2}{*}{225} & \multirow{2}{*}{78-634} & CI-opt & 2.6 & 0.5 & 4.9 & \multirow{2}{*}{0.0}\\ \cline{7-10} 
& & & & & & CI & 2.7 & 1 & 4.9 & \\ \cline{4-11}
& & & \multirow{2}{*}{38}&  \multirow{2}{*}{12} & \multirow{2}{*}{60-376} & CI-opt & 2.5 & 0.1 & 10.3 & \multirow{2}{*}{0.2}\\ \cline{7-10} 
& & & & & & CI & 2.7 & 1 & 10.5 & \\ \cline{4-11}
& & & \multirow{2}{*}{2-38}&  \multirow{2}{*}{1869} & \multirow{2}{*}{60-1238} & CI-opt & 3.4 & 8.1 & 5.6 & \multirow{2}{*}{0.1}\\ \cline{7-10} 
& & & & & & CI & 2.9 & 1 & 5.7 & \\ \Xcline{1-11}{1.5pt}
 \multicolumn{2}{|c|}{\multirow{16}{*}{ UMi SC}} & \multirow{6}{*}{LOS}& \multirow{2}{*}{28}&  \multirow{2}{*}{4} & \multirow{2}{*}{31-54} & CI-opt & 3.8 & 34.2 & 2.4 & \multirow{2}{*}{0.8}\\ \cline{7-10} 
 & & & & & & CI & 2.1 & 1 & 3.2 & \\ \cline{4-11}
 & & & \multirow{2}{*}{73}&  \multirow{2}{*}{6} & \multirow{2}{*}{27-54} & CI-opt & -0.7 & 46.6 & 3.9 & \multirow{2}{*}{1.2}\\ \cline{7-10} 
& & & & & & CI & 2.1 & 1 & 5.1 & \\ \cline{4-11}
& & & \multirow{2}{*}{28, 73}&  \multirow{2}{*}{10} & \multirow{2}{*}{27-54} & CI-opt & 0.8 & 50.0 & 4.3 & \multirow{2}{*}{0.1}\\ \cline{7-10} 
& & & & & & CI & 2.1 & 1 & 4.4 & \\ \cline{3-11}
& & \multirow{10}{*}{NLOS}& \multirow{2}{*}{2.9}&  \multirow{2}{*}{18} & \multirow{2}{*}{109-235} & CI-opt & 3.5 & 8.2 & 2.9 & \multirow{2}{*}{0.0}\\ \cline{7-10} 
 & & & & & & CI & 2.9 & 1 & 2.9 & \\ \cline{4-11}
&&& \multirow{2}{*}{28}&  \multirow{2}{*}{18} & \multirow{2}{*}{61-186} & CI-opt & 3.3 & 0.7 & 8.6 & \multirow{2}{*}{0.0}\\ \cline{7-10} 
 & & & & & & CI & 3.4 & 1 & 8.6 & \\ \cline{4-11}
 &&& \multirow{2}{*}{29}&  \multirow{2}{*}{16} & \multirow{2}{*}{109-235} & CI-opt & 3.6 & 5.0 & 4.9 & \multirow{2}{*}{0.0}\\ \cline{7-10} 
 & & & & & & CI & 3.1 & 1 & 4.9 & \\ \cline{4-11}
 & & & \multirow{2}{*}{73} & \multirow{2}{*}{30} & \multirow{2}{*}{48-190} & CI-opt & 2.9 & 0.1 & 7.4 & \multirow{2}{*}{0.0}\\ \cline{7-10} 
& & & & & & CI & 3.4 & 1 & 7.4 & \\ \cline{4-11}
& & & \multirow{2}{*}{2.9-73} &  \multirow{2}{*}{82} & \multirow{2}{*}{48-235} & CI-opt & 2.8 & 0.1 & 7.8 & \multirow{2}{*}{0.2}\\ \cline{7-10} 
& & & & & & CI & 3.2 & 1 & 8.0 & \\ \cline{1-11}
\end{tabular}
\end{center}
\end{table*}

\begin{table*}
\renewcommand{\arraystretch}{1.4}
\begin{center}
\caption{Parameters in the CI and CI-opt path loss models in the InH scenario. Freq. Range denotes frequency range. \# of Data Points represents the number of data points after distance binning and path loss thresholding. Dist. Range denotes distance range, CI-opt represents the CI model with an optimized free space reference distance $d_0$. $\Delta_\sigma$ denotes the difference in the SF standard deviation between the CI and CI-opt models.}~\label{tbl:InH_CI}
\begin{tabular}{|cc|c|c|c|c|c|c|c|c|c|}
\hline
 \multicolumn{2}{|c|}{Sce.} & Env.& \makecell{Freq.\\Range\\(GHz)} & \makecell{\# of\\Data \\Points} &\makecell{Dist. \\Range\\(m)}&Model & \makecell{$\PLE$} & $d_0 (m)$ & \makecell{$\sigma$ \\($\dB$)} & \makecell{$\Delta_\sigma$ \\($\dB$)} \\ \Xcline{1-11}{1.5pt}
 \multicolumn{2}{|c|}{\multirow{20}{*}{InH Office}} & \multirow{10}{*}{LOS}&  \multirow{2}{*}{2.9}&  \multirow{2}{*}{12} & \multirow{2}{*}{5-49} & CI-opt & 1.8 & 0.1 & 5.0 & \multirow{2}{*}{0.2}\\ \cline{7-10} 
 & & & & & & CI & 1.6 & 1 & 5.2 & \\ \cline{4-11}
&&& \multirow{2}{*}{28}&  \multirow{2}{*}{6} & \multirow{2}{*}{4-21} & CI-opt & 1.1 & 1.1 & 1.2 & \multirow{2}{*}{0.0}\\ \cline{7-10} 
 & & & & & & CI & 1.1 & 1 & 1.2 & \\ \cline{4-11}
 &&& \multirow{2}{*}{29}&  \multirow{2}{*}{12} & \multirow{2}{*}{5-49} & CI-opt & 1.5 & 0.9 & 4.5 & \multirow{2}{*}{0.0}\\ \cline{7-10} 
 & & & & & & CI & 1.5 & 1 & 4.5 & \\ \cline{4-11}
 & & & \multirow{2}{*}{73} & \multirow{2}{*}{6} & \multirow{2}{*}{4-21} & CI-opt & 0.4 & 3.7 & 1.2 & \multirow{2}{*}{1.8}\\ \cline{7-10} 
& & & & & & CI & 1.3 & 1 & 3.0 & \\ \cline{4-11}
& & & \multirow{2}{*}{2.9-73} &  \multirow{2}{*}{36} & \multirow{2}{*}{4-49} & CI-opt & 1.7 & 0.1 & 4.6 & \multirow{2}{*}{0.0}\\ \cline{7-10} 
& & & & & & CI & 1.5 & 1 & 4.6 & \\ \cline{3-11}
& & \multirow{10}{*}{NLOS}&  \multirow{2}{*}{2.9}&  \multirow{2}{*}{30} & \multirow{2}{*}{5-67} & CI-opt & 3.9 & 4.6 & 5.9 & \multirow{2}{*}{0.6}\\ \cline{7-10} 
 & & & & & & CI & 3.1 & 1 & 6.5 & \\ \cline{4-11}
&&& \multirow{2}{*}{28}&  \multirow{2}{*}{17} & \multirow{2}{*}{4-46} & CI-opt & 3.3 & 4.4 & 8.8 & \multirow{2}{*}{0.3}\\ \cline{7-10} 
 & & & & & & CI & 2.7 & 1 & 9.1 & \\ \cline{4-11}
 &&& \multirow{2}{*}{29}&  \multirow{2}{*}{29} & \multirow{2}{*}{5-67} & CI-opt & 4.4 & 4.7 & 6.4 & \multirow{2}{*}{0.8}\\ \cline{7-10} 
 & & & & & & CI & 3.3 & 1 & 7.2 & \\ \cline{4-11}
 & & & \multirow{2}{*}{73} & \multirow{2}{*}{15} & \multirow{2}{*}{4-42} & CI-opt & 2.8 & 0.5 & 9.1 & \multirow{2}{*}{0.1}\\ \cline{7-10} 
& & & & & & CI & 3.0 & 1 & 9.2 & \\ \cline{4-11}
& & & \multirow{2}{*}{2.9-73} &  \multirow{2}{*}{91} & \multirow{2}{*}{4-67} & CI-opt & 3.9 & 3.9 & 7.9 & \multirow{2}{*}{0.4}\\ \cline{7-10} 
& & & & & & CI & 3.1 & 1 & 8.3 & \\ \Xcline{1-11}{1.5pt}
 \multicolumn{2}{|c|}{\multirow{16}{*}{InH SM}} & \multirow{8}{*}{LOS}&  \multirow{2}{*}{2.9}&  \multirow{2}{*}{14} & \multirow{2}{*}{19-149} & CI-opt & 1.9 & 0.1 & 3.2 & \multirow{2}{*}{0.0}\\ \cline{7-10} 
 & & & & & & CI & 1.9 & 1 & 3.2 & \\ \cline{4-11}
 &&& \multirow{2}{*}{29}&  \multirow{2}{*}{14} & \multirow{2}{*}{19-149} & CI-opt & 1.8 & 7.6 & 3.1 & \multirow{2}{*}{0.0}\\ \cline{7-10} 
 & & & & & & CI & 1.9 & 1 & 3.1 & \\ \cline{4-11}
 & & & \multirow{2}{*}{61} & \multirow{2}{*}{14} & \multirow{2}{*}{19-149} & CI-opt & 1.6 & 50 & 3.4 & \multirow{2}{*}{0.0}\\ \cline{7-10} 
& & & & & & CI & 2.0 & 1 & 3.4 & \\ \cline{4-11}
& & & \multirow{2}{*}{2.9-61} &  \multirow{2}{*}{42} & \multirow{2}{*}{19-149} & CI-opt & 1.9 & 7.0 & 3.4 & \multirow{2}{*}{0.0}\\ \cline{7-10} 
& & & & & & CI & 1.9 & 1 & 3.4 & \\ \cline{3-11}
& & \multirow{8}{*}{NLOS}&  \multirow{2}{*}{2.9}&  \multirow{2}{*}{26} & \multirow{2}{*}{24-229} & CI-opt & 2.1 & 0.1 & 4.8 & \multirow{2}{*}{0.0}\\ \cline{7-10} 
 & & & & & & CI & 2.2 & 1 & 4.8 & \\ \cline{4-11}
 &&& \multirow{2}{*}{29}&  \multirow{2}{*}{26} & \multirow{2}{*}{24-229} & CI-opt & 2.2 & 0.1 & 4.2 & \multirow{2}{*}{0.0}\\ \cline{7-10} 
 & & & & & & CI & 2.3 & 1 & 4.2 & \\ \cline{4-11}
 & & & \multirow{2}{*}{61} & \multirow{2}{*}{26} & \multirow{2}{*}{24-229} & CI-opt & 2.3 & 0.1 & 4.5 & \multirow{2}{*}{0.0}\\ \cline{7-10} 
& & & & & & CI & 2.5 & 1 & 4.5 & \\ \cline{4-11}
& & & \multirow{2}{*}{2.9-61} &  \multirow{2}{*}{78} & \multirow{2}{*}{24-229} & CI-opt & 2.2 & 0.1 & 4.8 & \multirow{2}{*}{0.0}\\ \cline{7-10} 
& & & & & & CI & 2.3 & 1 & 4.8 & \\ \cline{1-11}
\end{tabular}
\end{center}
\end{table*} 

One may argue that a close-in reference distance other than 1 m may be a better approach to maximize model accuracy of the CI model\cite{Abdallah2014Further,Alv03}. Some of the authors of this paper, in fact, originally used $d_0$ values greater than 1 m in past research in order to ensure the model would only be used in the far field of directional antennas\cite{Ben11,Rap_Ben12,Rap13:Access}, but they later found a 1 m reference was more suitable for use as a standard, due to the fact that there was very little difference in standard deviation when using a 1 m reference distance (i.e., model error was not significantly different when using a different value of $d_0$\cite{Rap15:TCOM}), and given the fact that very few or any users will be within the first few meters of the transmitter antenna.

To compare the performance of the CI model between using a 1 m free space reference distance and an optimized or empirically determined free space reference distance $d_0$, as proposed in\cite{Abdallah2014Further,Alv03}, we used the 30 measurement data sets from Nokia/AAU, UT, NYU, and Qualcomm to compare model parameters and standard deviations. Tables~\ref{tbl:UMa_CI} and~\ref{tbl:InH_CI} list the model parameters in the 1 m CI model as compared to the CI model with an optimized $d_0$ (CI-opt) at various frequencies ranging from 2 GHz to 73 GHz for the UMa, UMi, and InH scenarios in both LOS and NLOS environments, where the PLE and $d_0$ for CI-opt were jointly optimized via the MMSE method demonstrated in the Appendix (to preclude unreasonable $d_0$ values caused by the sparsity of some data sets, the range of $d_0$ was set to between 0.1 m and 50 m). All of the scattered path loss data samples were locally averaged over 2 m distance bins (other binning values can also be explored, and we found little difference in results using 2, 5, or 10 m local average bins), in order to remove the small-scale fading effects and to reduce the difference in the number of data points across measurement campaigns. In addition, all path loss values weaker than FSPL at 1 m plus 100 dB were not considered for analysis, based on the reasonable assumption that there would be fewer weaker measurements at higher frequencies due to the greater path loss in the first meter, so a frequency-dependent signal threshold was implemented to ensure that the measured data sets would slightly emphasize more measurements at the higher frequencies, resulting in a relatively comparable number of points for the different frequencies from various measurement campaigns. We note that the results of this paper were not heavily influenced by the binning or frequency-dependent thresholding, but these approaches were found to yield comparable coverage distances over the multiple frequencies based on the particular antennas and transmit powers used.

\begin{figure}
\centering
 \includegraphics[width=3.4in]{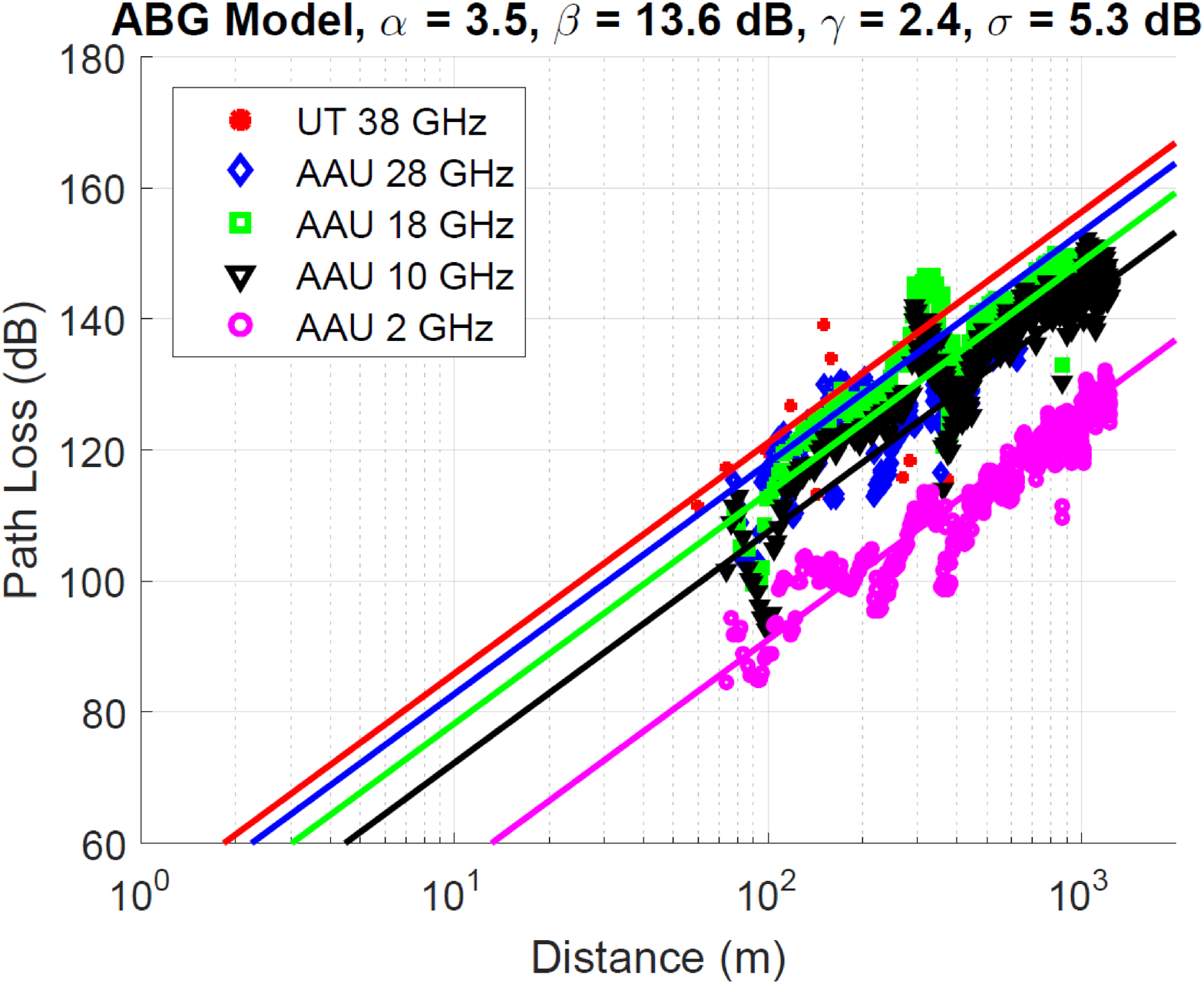}
    \caption{ABG path loss model in the UMa scenario across different frequencies and distances in the NLOS environment. Model parameters using all of the displayed data are given at the top of the graph.}
    \label{fig:UMi_SC_NLOS_ABG}
\end{figure}
\begin{figure}
\centering
 \includegraphics[width=3.4in]{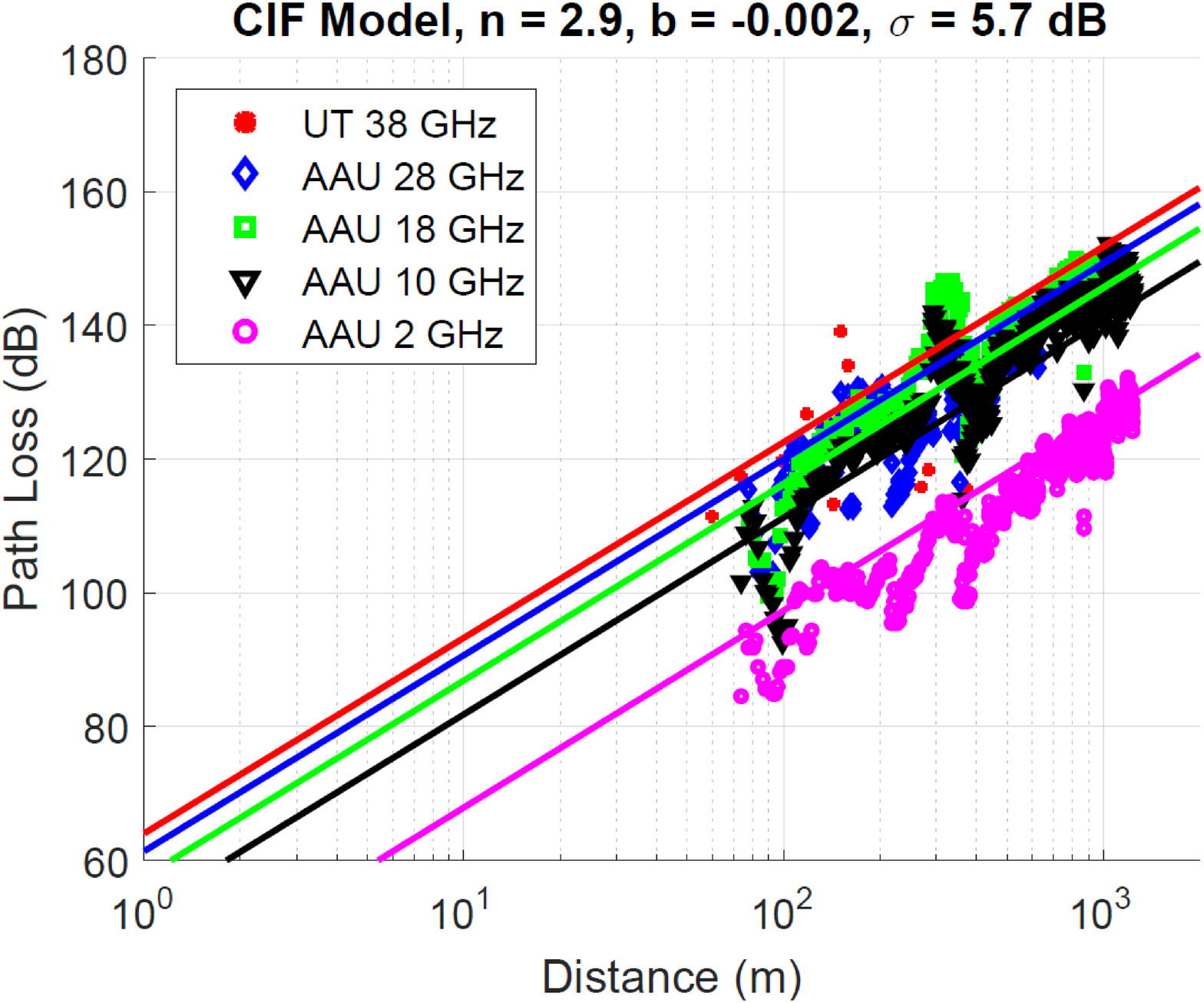}
    \caption{CIF path loss model in the UMa scenario across different frequencies and distances in the NLOS environment. Model parameters using all of the displayed data are given at the top of the graph.}
    \label{fig:UMi_SC_NLOS_CIF}
\end{figure}
\begin{figure}
\centering
 \includegraphics[width=3.4in]{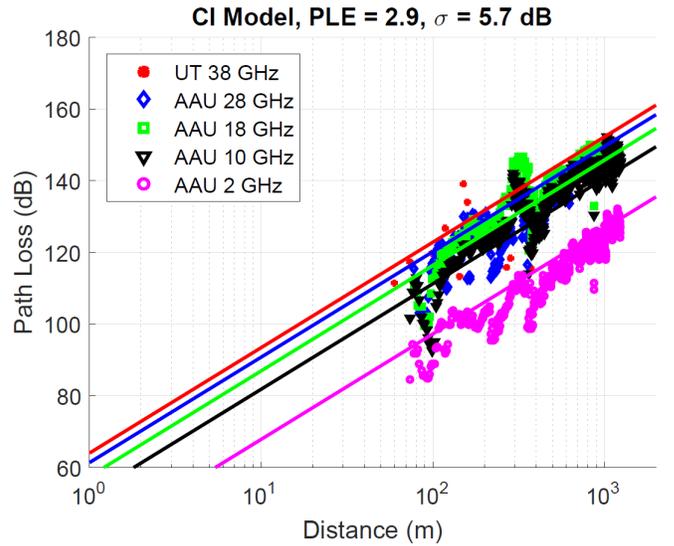}
    \caption{CI path loss model in the UMa scenario across different frequencies and distances in the NLOS environment. Model parameters using all of the displayed data are given at the top of the graph.}
    \label{fig:UMi_SC_NLOS_CI}
\end{figure}

As shown in Tables~\ref{tbl:UMa_CI} and~\ref{tbl:InH_CI}, for both outdoor and indoor scenarios, the SF between using $d_0$ = 1 m and an optimized $d_0$ differs by no more than 0.3 dB in most cases (more than an order of magnitude smaller than the standard deviation). Note that the only significant differences in error between the $d_0$ = 1  m and the optimized $d_0$ value occur when there are very few measurement points, and the PLE in CI-opt generally has a physically unreasonable value in these rare cases (e.g., the PLE is less than 1, indicating much less loss than a metal waveguide; or the PLE is negative, indicating decrease of path loss with distance; or the PLE is unreasonably high). For the majority of the measurement sets, the 1 m free space reference distance model $\chi_{\sigma}$ is always within 0.1 dB of the optimized $d_0$ model, illustrating virtually no difference in standard deviation between the two approaches. Therefore, the 1 m CI model provides sufficiently accurate fitting results compared to the CI-opt model, and requires only one model parameter (PLE) to be optimized by the adoption of a 1 m standard close-in free space reference distance, while the CI-opt model requires two model parameters (PLE and $d_0$) for modeling the mean path loss over distance, and sometimes yields unrealistic PLEs and reference distances. For the remainder of this paper, the CI model~\eqref{CI1} is assumed to use $d_0$ = 1 m, as suggested in\cite{Rap15:TCOM}.

The ABG~\eqref{ABG1}, CI~\eqref{CI1} and CIF~\eqref{CIF1} models with $d_0=1$ m are a function of both distance and frequency, where the CI and CIF models have frequency dependence expressed primarily by the frequency-dependent FSPL term~\eqref{FSPL} in the first meter of free space propagation. While the ABG model offers some physical basis in the $\alpha$ term, being based on a 1 m reference distance similar to the $n$ term in~\eqref{CI1} and~\eqref{CIF1}, it departs from physics when introducing both an offset $\beta$ (which is an optimization parameter that is not physically based), and a frequency weighting term $\gamma$ that has no proven physical basis, although recent measurements show that the path loss increases with frequency across the mmWave band in the indoor office scenario\cite{Deng15} (both the $\beta$ and $\gamma$ parameters are used for curve fitting, as was done in the WINNER floating-intercept (AB) model)\cite{WINNER,Rap15:TCOM,GRM13:Globecom}. It is noteworthy that the ABG model is identical to the CI model if we equate $\alpha$ in the ABG model in~\eqref{ABG1} with the PLE $n$ in the CI model in~\eqref{CI1}, $\gamma$ in~\eqref{ABG1} with the free space PLE of 2, and $\beta$ in~\eqref{ABG1} with $20log_{10}(4\pi\times 10^9/c)$  in~\eqref{FSPL}. 

Using the three path loss models described above, and the 30 measurement data sets over a wide range of microwave and mmWave frequencies (2 to 73 GHz) and distances (4 to 1238 m), we computed the path loss model parameters for the three models. The PLE in the CI model, the $n$ and $b$ in the CIF model, and the $\alpha$, $\beta$, and $\gamma$ parameters in the ABG model were all calculated via the MMSE fit on all of the path loss data from all measured frequencies and distances for a given scenario (UMa, UMi, or InH), using closed-form solutions that minimize the SF standard deviation, as detailed in the Appendix. In order to focus solely on the comparison of propagation models, we separated LOS and NLOS measurements, and did not include the probabilities of LOS or NLOS, although we note that such probability models as described in\cite{Akd14,Samimi15:WCL,Rap15:TCOM,3GPP:25996,3GPP:36814,Bai15} may exploit the results of this work. 

Figs.~\ref{fig:UMi_SC_NLOS_ABG} -~\ref{fig:UMi_SC_NLOS_CI} show scatter plots of all the data sets optimized for the ABG, CIF, and CI models in the UMa scenario in the NLOS environment, respectively. Table~\ref{tbl:UMi_UMa} summarizes the path loss parameters in the ABG, CI, and CIF models for the UMa, UMi, and InH scenarios in both LOS and NLOS environments. As shown in Table~\ref{tbl:UMi_UMa}, the CI and CIF models each provide a PLE of 2.0, 2.1, 1.5, and 1.9  in the LOS environment for the UMa, UMi SC, InH office and InH SM scenarios, respectively, which agrees well with a free space PLE of 2 in UMa, UMi SC, and InH SM settings, or models the waveguiding effects in the InH office scenario, respectively. Although the CI and CIF models yield slightly higher SF standard deviation than the ABG model in most cases, this increased standard deviation is usually a fraction of a dB and is within standard measurement error arising from frequency and temperature drift, connector and cable flex variations, and calibration errors in an actual measurement campaign. Notably, these errors are often an order of magnitude less than the corresponding actual SF standard deviations in all three models. It is noteworthy that the CIF model even renders lower SF standard deviations than the ABG model for the LOS InH office and NLOS InH SM scenarios, indicating the greater accuracy of CIF compared to ABG in these settings, even though the CIF model has fewer optimization parameters. Furthermore, for the UMa and LOS UMi SC scenarios, the CI and CIF models always yield identical PLEs and standard deviations for the same data set, and the $b$ parameter in the CIF model is virtually zero. For the NLOS UMi SC, and InH SM scenarios, $b$ in the CIF model is slightly positive, implying that path loss increases with frequency beyond the first meter of free space propagation. 

\begin{table*}
\renewcommand{\arraystretch}{1.4}
\begin{center}
\caption{Parameters in the ABG, CI, and CIF path loss models in UMa, UMi, and InH scenarios (Sce.) in both LOS and NLOS environments (Env.). Freq. Range denotes frequency range, and Dist. Range denotes distance range. \# of Data Points represents the number of data points after distance binning and path loss thresholding. $\Delta_\sigma$ denotes the difference in the SF standard deviation between the CI or CIF model and the ABG model.}~\label{tbl:UMi_UMa}
\begin{tabular}{|c|c|c|c|c|c|c|c|c|c|c|}
\hline 
 Sce. & Env.& \makecell{Freq.\\Range\\(GHz)}& \makecell{\# of\\Data Points}&\makecell{Dist. \\Range\\(m)}&Model & \makecell{$\alpha$ (ABG)\\or $\PLE$ (CI)\\or $n$ (CIF)} & \makecell{$\beta$ (ABG)\\($\dB$)} & \makecell{$\gamma$ (ABG)\\or $b$ (CIF)} & \makecell{$\sigma$ \\($\dB$)} & \makecell{$\Delta_\sigma$ \\($\dB$)} \\ \Xcline{1-11}{1.5pt}
\multirow{6}{*}{\makecell{UMa}} & \multirow{3}{*}{LOS}& \multirow{3}{*}{2-38}& \multirow{3}{*}{1032}& \multirow{3}{*}{60-930} & ABG & 1.9 & 35.8 & 1.9 & 2.4 & - \\ \cline{6-11}
 & & & & & CI & 2.0 & - & - & 2.4 & 0.0 \\ \cline{6-11}
 & & & & & CIF & 2.0 & - & -0.014 & 2.4 & 0.0 \\ \cline{2-11}
 & \multirow{3}{*}{NLOS}  & \multirow{3}{*}{2-38}& \multirow{3}{*}{1869}  & \multirow{3}{*}{61-1238} & ABG & 3.5 & 13.6 & 2.4 & 5.3 & - \\ \cline{6-11}
  & & & & & CI & 2.9 & - & - & 5.7 & 0.4 \\ \cline{6-11}
  & & & & & CIF & 2.9 & - & -0.002 & 5.7 & 0.4 \\ \Xcline{1-11}{1.5pt}
  \multirow{6}{*}{\makecell{UMi SC}} & \multirow{3}{*}{LOS}& \multirow{3}{*}{28, 73}& \multirow{3}{*}{10}& \multirow{3}{*}{27-54} & ABG & 1.1 & 46.8 & 2.1 & 4.3 & - \\ \cline{6-11}
 & & & & & CI & 2.1 & - & - & 4.4 & 0.1 \\ \cline{6-11}
  & & & & & CIF & 2.1 & - & 0.003 & 4.4 & 0.1 \\ \cline{2-11}
 & \multirow{3}{*}{NLOS}  & \multirow{3}{*}{2.9-73}& \multirow{3}{*}{82}  & \multirow{3}{*}{48-235} & ABG & 2.8 & 31.4 & 2.7 & 6.8 & - \\ \cline{6-11}
  & & & & & CI & 3.2 & - & - & 8.0 & 1.2 \\ \cline{6-11}
  & & & & & CIF & 3.2 & - & 0.076 & 7.1 & 0.3 \\ \Xcline{1-11}{1.5pt}
  \multirow{6}{*}{\makecell{InH Office}} & \multirow{3}{*}{LOS}& \multirow{3}{*}{2.9-73}& \multirow{3}{*}{36}& \multirow{3}{*}{4-49} & ABG & 1.6 & 32.9 & 1.8 & 4.5 & - \\ \cline{6-11}
 & & & & & CI & 1.5 & - & - & 4.6 & 0.1 \\ \cline{6-11}
 & & & & & CIF & 1.5 & - & -0.102 & 4.4 & -0.1 \\ \cline{2-11}
 & \multirow{3}{*}{NLOS}  & \multirow{3}{*}{2.9-73}& \multirow{3}{*}{91}  & \multirow{3}{*}{4-67} & ABG & 3.9 & 19.0 & 2.1 & 7.9 & - \\ \cline{6-11}
  & & & & & CI & 3.1 & - & - & 8.3 & 0.4 \\ \cline{6-11}
  & & & & & CIF & 3.1 & - & -0.001 & 8.3 & 0.4 \\ \Xcline{1-11}{1.5pt}
   \multirow{6}{*}{\makecell{InH SM}} & \multirow{3}{*}{LOS}& \multirow{3}{*}{2.9-61}& \multirow{3}{*}{42}& \multirow{3}{*}{19-149} & ABG & 1.9 & 31.2 & 2.2 & 3.3 & - \\ \cline{6-11}
 & & & & & CI & 1.9 & - & - & 3.4 & 0.1 \\ \cline{6-11}
 & & & & & CIF & 1.9 & - & 0.042 & 3.3 & 0.0 \\ \cline{2-11}
 & \multirow{3}{*}{NLOS}  & \multirow{3}{*}{2.9-61}& \multirow{3}{*}{78}  & \multirow{3}{*}{24-229} & ABG & 2.0 & 34.4 & 2.3 & 4.6 & - \\ \cline{6-11}
  & & & & & CI & 2.3 & - & - & 4.8 & 0.2 \\ \cline{6-11}
  & & & & & CIF & 2.3 & - & 0.054 & 4.5 & -0.1 \\ \cline{1-11}
\end{tabular}
\end{center}
\end{table*} 

\begin{table*}
\renewcommand{\arraystretch}{1.4}
\begin{center}
\caption{Parameters in the AB/ABG and CI (i.e., CIF when $b$ = 0) path loss models in the UMa and UMi scenarios (Sce.) in the NLOS environment (Env.) for different frequency (Freq.) and distance (Dist.) ranges. \# of Data Points represents the number of data points after distance binning and path loss thresholding.}~\label{tbl:UMa}
\begin{tabu}{|c|c|c|c|c||c|[1.7pt black]c|c|c|[1.7pt black]c|[1.7pt black]c|[1.7pt black]c|}
\hline 
 Sce. & Env.&\makecell{Freq./Freq. \\Range (GHz)} & \makecell{\# of\\Data Points}&\makecell{Dist. Range\\(m)} & $n^{CI}$& $\alpha^{ABG}$ & \makecell{$\beta^{ABG}$ \\($\dB$)} & $\gamma^{ABG}$ &\makecell{$\sigma^{CI}$ \\($\dB$)} & \makecell{$\sigma^{ABG}$ \\($\dB$)} & \makecell{$\sigma^{CI}-\sigma^{ABG}$ \\($\dB$)}\\ \cline{1-12} \hline\hline
\multirow{5}{*}{\makecell{UMa}} & \multirow{5}{*}{NLOS} & 2 & 583 & 74-1238 & 2.8 & 3.3 & 19.6 & 2 & 3.5 & 3.2 & 0.3\\ \cline{3-12}
 & & 18 & 468 &78-1032 & 3.0 & 3.2 & 28.5 & 2 & 4.5 & 4.4 & 0.1\\ \cline{3-12}
 & & 28 & 225 & 78-634 & 2.7 & 2.6 & 34.0 & 2 & 4.9 & 4.9 & 0.0\\ \cline{3-12}
& & 38 & 12 & 60-376 & 2.7 & 1.0 & 69.3 & 2 & 10.5 & 9.6 & 0.9\\ \tabucline[1.4pt black off 0pt]{3-12}
& & 2-38 & 1869 & 60-1238 & 2.9 & 3.5 & 13.6 & 2.4 & 5.7 & 5.3 & 0.4\\ \cline{1-12} \hline\hline
\multirow{5}{*}{\makecell{UMi SC}} & \multirow{5}{*}{NLOS} & 2.9 & 18 & 109-235 & 2.9 & 3.5 & 18.9 & 2 & 2.9 & 2.9 & 0.0\\ \cline{3-12}
&& 28 & 18 & 61-186 & 3.4 & 3.3 & 34.1 & 2 & 8.6 & 8.6 & 0.0\\ \cline{3-12}
& & 29 & 16 & 109-235 & 3.1 & 3.6 & 21.3 & 2 & 4.9 & 4.9 & 0.0\\ \cline{3-12}
& & 73 & 30 & 48-190 & 3.4 & 2.9 & 42.6 & 2 & 7.4 & 7.4 & 0.0\\ \tabucline[1.4pt black off 0pt]{3-12}
& & 2.9-73 & 82 & 48-235 & 3.2 & 2.8 & 31.4 & 2.7 & 8.0 & 6.8 & 1.2\\ \cline{1-12} \hline
\end{tabu}
\end{center}
\end{table*} 

\begin{figure}
\centering
 \includegraphics[width=3.4in]{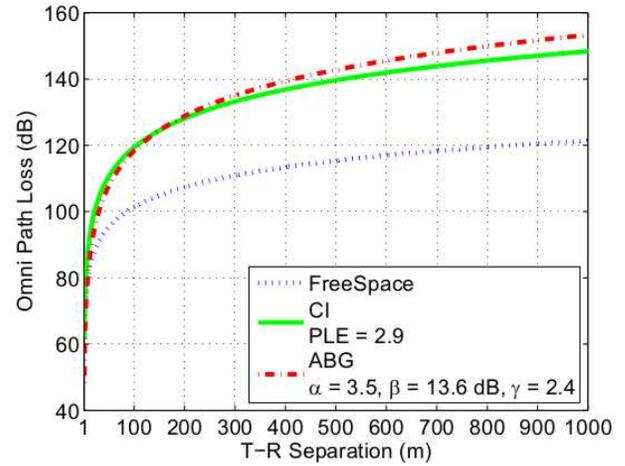}
    \caption{Example comparison of free space, CI and ABG path loss models at 28 GHz for the UMa NLOS environment using the parameters derived with measurements from 2 - 38 GHz in Table~\ref{tbl:UMi_UMa}. Note how the ABG model estimates 5 dB less signal power (i.e., 5 dB less out-of-cell interference) at 1 km when compared to CI.}
    \label{fig:FS_CI_ABG}
\end{figure}

Table~\ref{tbl:UMa} lists the model parameters in the ABG and CI models at different frequencies in the NLOS environment for the UMa and UMi scenarios, with the last line for each scenario showing the parameters for the multi-frequency model. Note that for single frequencies, $\gamma$ in the ABG model is set to 2, thus reverting to the AB model used in 3GPP and WINNER II channel models\cite{WINNER,3GPP:25996,3GPP:36814}, and the CIF model reverts to the CI model. Fig.~\ref{fig:FS_CI_ABG} illustrates a useful example of the CI and ABG models as compared to ideal free space path loss at 28 GHz for the UMa NLOS environment, using the parameters for 2 - 38 GHz in Table~\ref{tbl:UMa}. Fig.~\ref{fig:FS_CI_ABG} is useful since it shows how any one of the three path loss models might be used at a particular single frequency in wireless system design, after the multi-frequency model had been developed using a wide range of data over a vast range of frequencies (in this case, the four measurement data sets for the UMa scenario listed in Table~\ref{tbl:UMa}). 

A few key observations can be obtained from these figures and Table~\ref{tbl:UMa}. First, the $\alpha$ and $\beta$ parameters in the AB model can vary as widely as 2.3 and 49.7 dB across frequencies, respectively, as shown in Table~\ref{tbl:UMa}. The large variation of $\alpha$ and $\beta$ in the AB model was also observed in\cite{Rap15:TCOM}. Second, the PLE $n$ in the CI model varies only marginally for the single frequency case, with a largest variation of merely 0.5 for all the scenarios. The SF standard deviations for the CI and ABG models differ by only a fraction of a dB over all frequencies and distances in most cases, and the difference is less than an order of magnitude of the SF for either model, making the models virtually identical in accuracy over frequency and distance. There is a case for UMi where the ABG model has 1.2 dB lower SF standard deviation than the CI model, but there are only 82 data points in this case, and recent working using a much larger data set showed only 0.4 dB difference (8.2 dB for CI and 7.8 dB for ABG) for the UMi SC NLOS scenario\cite{Han:VTC}, and this difference is more than an order of magnitude smaller than either standard deviation. 

As shown in Fig.~\ref{fig:FS_CI_ABG}, the parameters derived from 2 to 38 GHz for the UMa NLOS environment, when applied at 28 GHz, indicate that the ABG NLOS model underestimates path loss to be \textit{much less than free space} when very close to the transmitter (a nonsensical result!) and predicts much less path loss than CI NLOS out to $\sim$ 30 m. Perhaps more importantly, the floating-intercept ABG model overestimates path loss (i.e., underestimates interference) at greater distances compared with the CI model at far distances\cite{Rap15:TCOM}. These results are clearly seen by comparing the path loss vs. distance end-points in Figs.~\ref{fig:UMi_SC_NLOS_ABG},~\ref{fig:UMi_SC_NLOS_CI} and~\ref{fig:FS_CI_ABG}. The CI model is thus more conservative when analyzing interference-limited systems at larger distances and more realistic when modeling NLOS signal strengths at close-in distances.

From the above analysis, the CI model provides more stability and intrinsic accuracy at distance end-points using fewer parameters (i.e., PLE and $\chi_{\sigma}^{\CI}$) across wide ranges of frequencies with only a fraction of a decibel higher SF standard deviation in most cases when compared to the four-parameter ABG model. The CI model is anchored to FSPL in the first meter, and gives intuitive meaning through the PLE parameter, since 10$n$ mathematically describes the path loss in dB with respect to a decade increase of distance beginning at 1 m, making it very easy to compute power or path loss over distance. Only a very subtle change of a single constant is needed to the AB/ABG model to implement the simpler CI/CIF model, i.e., replacing the floating intercept parameter with a FSPL term that is physically based and is inherently a function of frequency. While Tables~\ref{tbl:UMi_UMa} and~\ref{tbl:UMa} show how the ABG, CI, and CIF models all provide comparable curve fitting standard deviations over a wide frequency range, we now show that the CI and CIF models offer superior accuracy and reliability when subject to extensive sensitivity analyses. 

\section{Prediction and Sensitivity Performance}
This section investigates the prediction accuracy and sensitivity of the three path loss models, i.e., ABG, CI and CIF. Because of the vast number of experimental data points provided by the authors, it was possible to test the efficacy of the path loss models in situations where they are used outside of the particular frequencies, locations, or distances. Prediction performance and model sensitivity were tested by creating path loss models using a subset of the measurements (to obtain the optimized model parameters) and then testing those resulting models against the other subset of measurements (which were outside of the data sets used to generate the original model parameters). This test is needed to establish whether engineers could use the models with confidence in new scenarios or distances or frequencies different than what were used to form the original models. If future systems use more transmit power or have greater range than the measurement systems used to derive the model parameters, or are to be used at different frequencies than what were measured to produce the models, a sensitivity analysis such as this is critical for comparing and selecting path loss models.

The measured data from all experiments for the UMa, UMi SC, and InH office scenarios shown in Table~\ref{tbl:UMi_UMa} are split into two sets: a measurement set and a prediction set, where the term \textit{measurement set} refers to the set of measured data used to compute the optimum (i.e., minimum SF standard deviation) parameters of the path loss model, and the term \textit{prediction set} refers to a different set of measured data that is scattered about the distance-dependent mean path loss model constructed from the measurement set. For a specific path loss model (e.g., ABG, CI, or CIF), the SF standard deviation is calculated using the measured data in the prediction set as distributed about the distance-dependent mean path loss model constructed from the measurement set. As the measurement set varies with distance, frequency, or city, as explained below, the optimized model parameters computed from the measurement set, as well as the SF standard deviation for the prediction set (i.e., the prediction error), also change. Therefore, two types of comparisons are simultaneously performed as the measurement set varies: first, the SF standard deviation for the prediction set about the model formed from the measurement set is computed and compared for each of the three path loss models in order to compare the accuracy for each model under identical measurement set conditions; second, the optimized model parameters from the measurement set are determined and compared between the three path loss models, to determine the sensitivity and stability  of the model parameters over different sets of measurement data. Only the NLOS data are used in this prediction performance and sensitivity study, since NLOS environments offer greater variability, higher SF standard deviation, and are most likely to produce errors in 5G analysis and simulation.

\subsection{Prediction in Distance}
In this subsection, the total data set of each of the UMa, UMi SC, and InH office NLOS data of Table~\ref{tbl:UMi_UMa} is used and broken up into a measurement set and a prediction set based on distance. The prediction set was kept fixed in this investigation and the measurement sets were varied over distance, where the optimum model parameters (corresponding to the minimum SF standard deviation) were computed for each specific measurement set. The measurement sets included measured data at distances which kept getting further away from the prediction set.

\begin{figure}
\centering
 \includegraphics[width=3.4in]{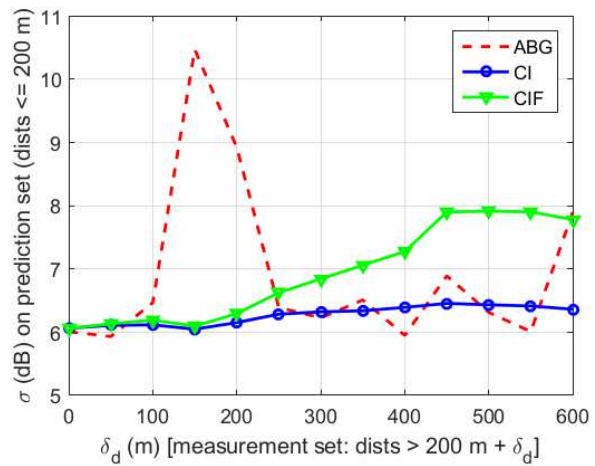}
    \caption{Shadow fading standard deviation of the ABG, CI, and CIF path loss models for prediction in distance when the prediction set is close to the transmitter in the UMa scenario.}
    \label{fig:Pre_Dist_larger_UMa}
\end{figure}

\begin{figure}
\centering
 \includegraphics[width=3.4in]{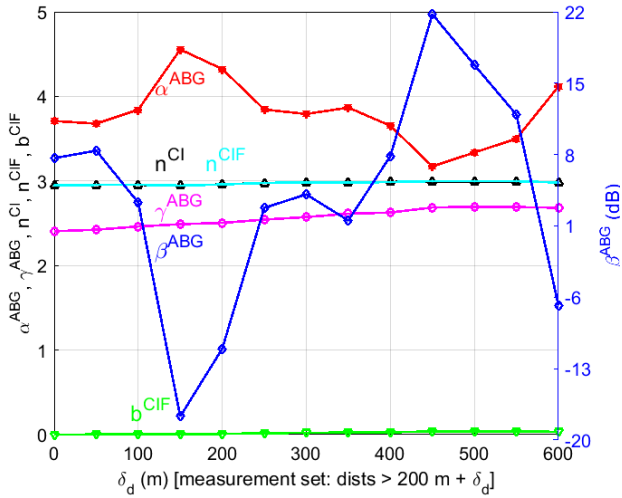}
    \caption{Parameters of the ABG, CI, and CIF path loss models for prediction in distance when the prediction set is close to the transmitter in the UMa scenario. Note that the scale for $\beta$ (dB) in the ABG model is to the right.}
    \label{fig:Pre_Dist_larger_para_UMa}
\end{figure}

The first investigation of this experiment is for the case when the prediction set contains measurement points that are closer to the TX (base station) than the measurement set. In this case, the prediction set is all the measured data with distances smaller than or equal to $d_{max}$ = 200 m, and the measurement sets varied to include all distances greater than $d_{max}+\delta_d~(\delta_d \geqslant 0)$. Fig.~\ref{fig:Pre_Dist_larger_UMa} and Fig.~\ref{fig:Pre_Dist_larger_para_UMa} show the prediction errors and parameter variations of the ABG, CI, and CIF models for prediction in distance in the UMa scenario. As can be seen in Fig.~\ref{fig:Pre_Dist_larger_UMa}, the prediction error of the CIF model generally increases with the increase of the distance between the two data sets. However, remarkably, the CI path loss model has a constant SF standard deviation for the prediction set, regardless of how far away the measurement set gets. On the other hand, the SF standard deviation of the ABG model over the prediction set varies substantially as $\delta_d$ increases. For the CI model, the largest difference in the standard deviation of the scattered data in the prediction set, around the optimized model derived from the measurement set, is only 0.4 dB across the entire range of $\delta_d$ (from 0 to 600 m), and about 2 dB for the CIF model, while the standard deviation of the ABG model reaches as high as 10.5 dB when $\delta_d$ = 150 m, and varies by 4.5 dB across the entire range of $\delta_d$. This shows how erratic and sensitive the ABG model is to the particular data used to create the model parameters, and illustrates the heightened sensitivity for certain situations when using the ABG model --- no such problems exist for the CI or CIF model. The parameter stability of the PLE in the CI model and the $n$ and $b$ values in the CIF model is much better than the parameters of the ABG model when varying the distance between the two sets, as seen in Fig.~\ref{fig:Pre_Dist_larger_para_UMa}. In particular, the $\alpha$ of the ABG model can vary a lot (3.2 to 4.6), which could have significant effects in system-level simulations, as the level of signal strength or interference greatly depends on the value of $\alpha$ (i.e., the distance-related parameter). In addition, the $\beta$ of the ABG model can vary by 39.5 dB.

\begin{figure}
\centering
 \includegraphics[width=3.4in]{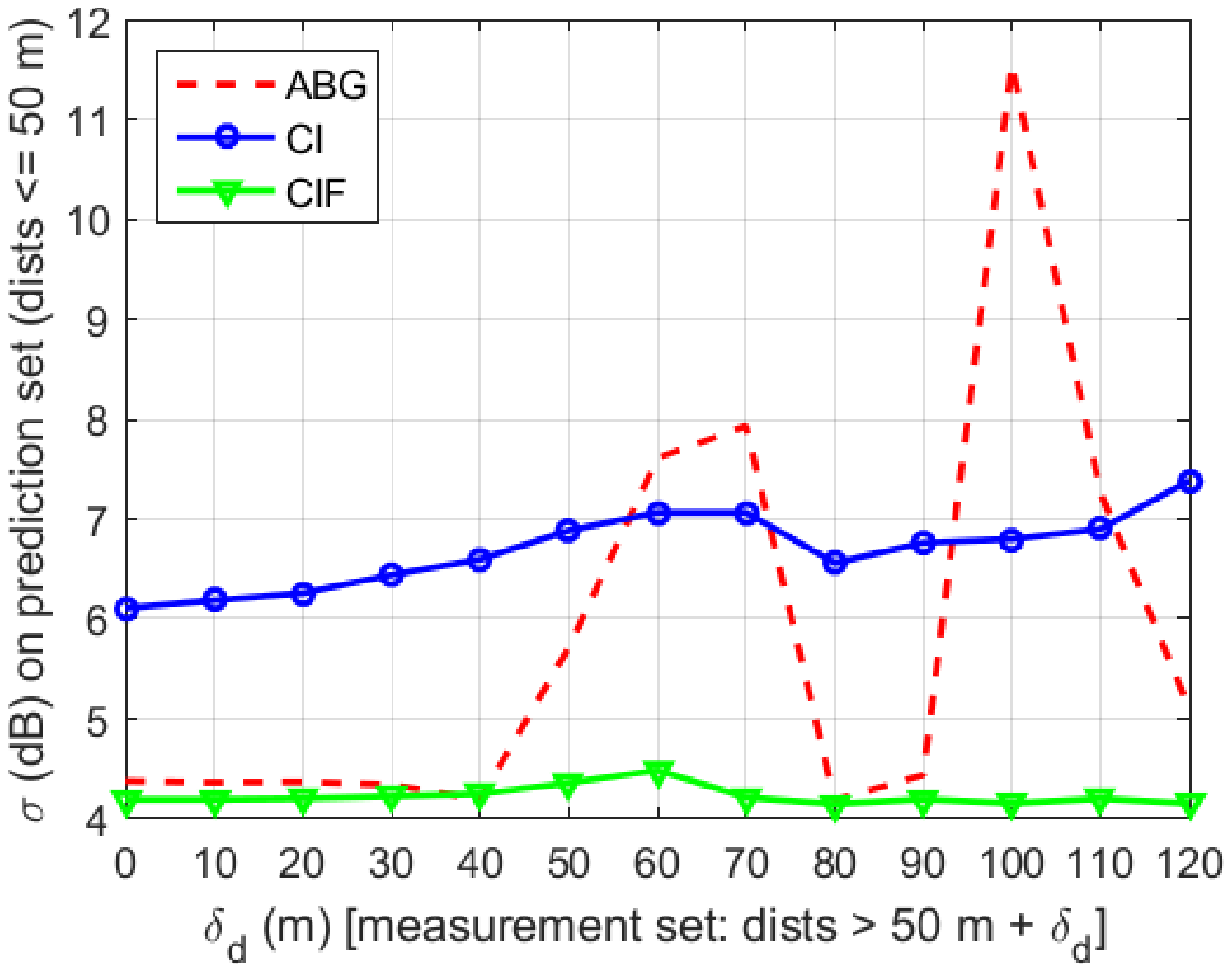}
    \caption{Shadow fading standard deviation of the ABG, CI, and CIF path loss models for prediction in distance when the prediction set is close to the transmitter in the UMi SC scenario.}
    \label{fig:Pre_Dist_larger_UMi}
\end{figure}

\begin{figure}
\centering
 \includegraphics[width=3.4in]{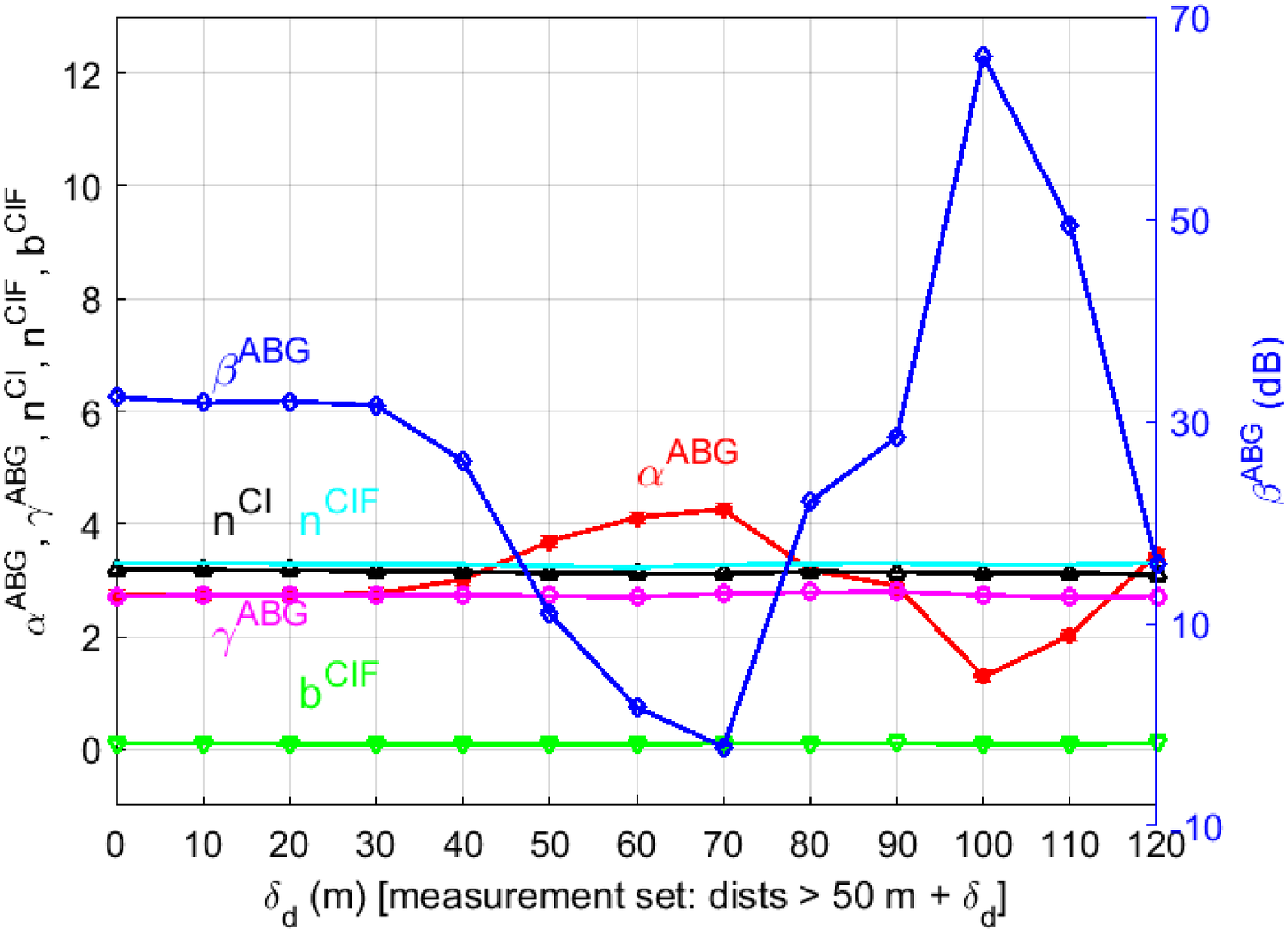}
    \caption{Parameters of the ABG, CI, and CIF path loss models for prediction in distance when the prediction set is close to the transmitter in the UMi SC scenario. Note that the scale for $\beta$ (dB) in the ABG model is to the right.}
    \label{fig:Pre_Dist_larger_para_UMi}
\end{figure}

\begin{figure}
\centering
 \includegraphics[width=3.4in]{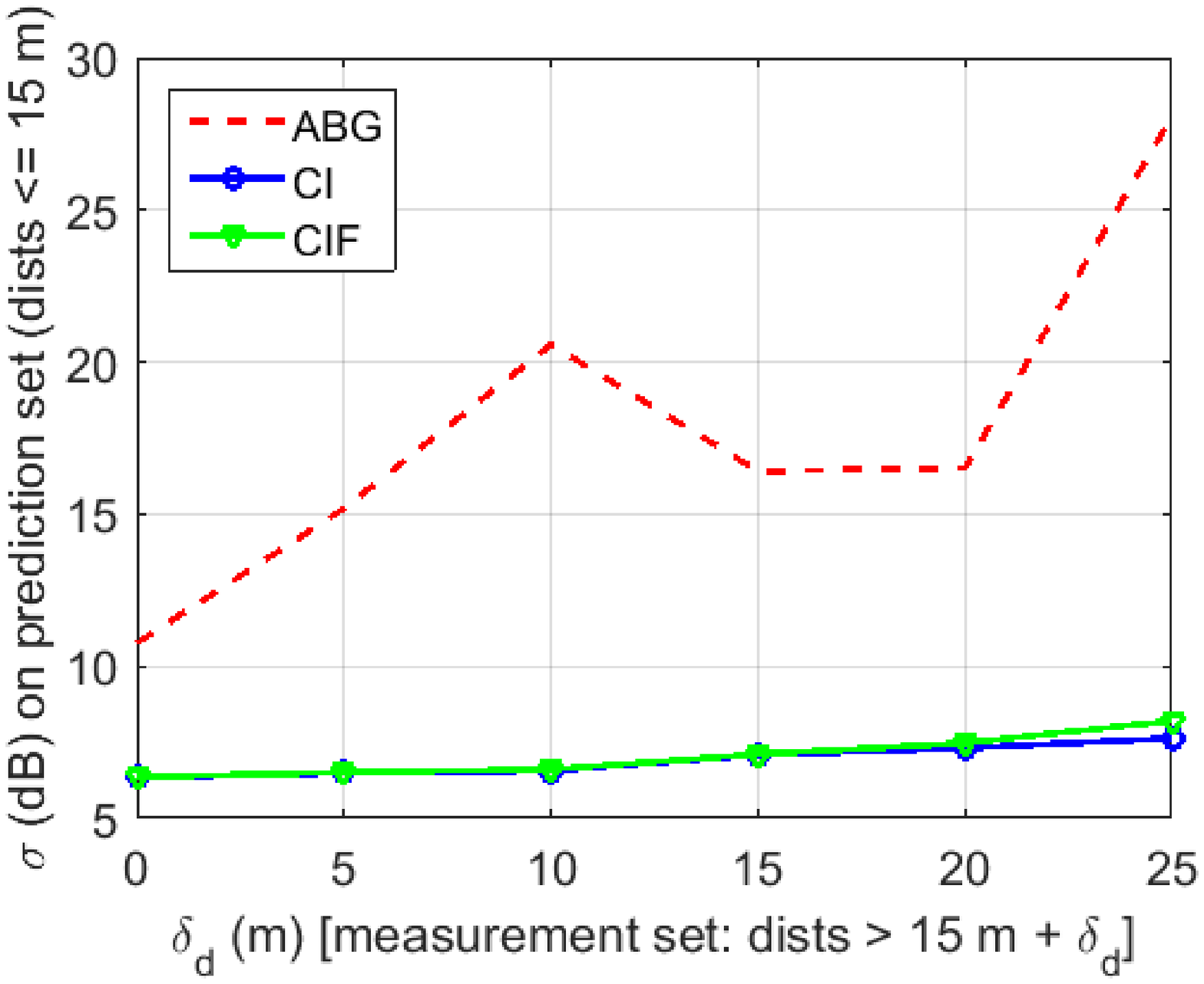}
    \caption{Shadow fading standard deviation of the ABG, CI, and CIF path loss models for prediction in distance when the prediction set is close to the transmitter in the InH office scenario.}
    \label{fig:Pre_Dist_larger_InH}
\end{figure}

\begin{figure}
\centering
 \includegraphics[width=3.4in]{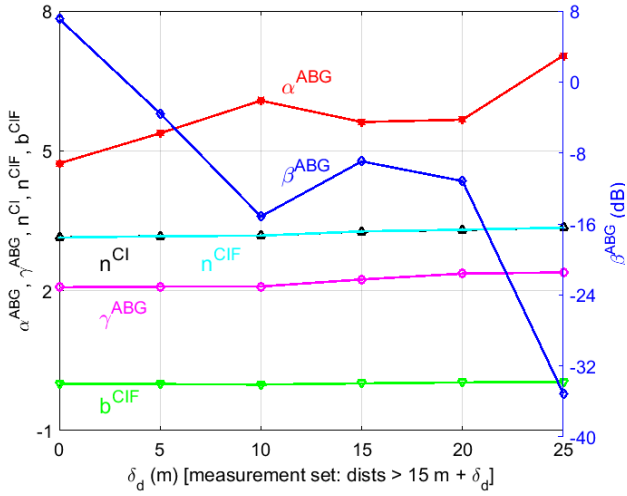}
    \caption{Parameters of the ABG, CI, and CIF path loss models for prediction in distance when the prediction set is close to the transmitter in the InH office scenario. Note that the scale for $\beta$ (dB) in the ABG model is to the right.}
    \label{fig:Pre_Dist_larger_para_InH}
\end{figure}

For the UMi scenario, the prediction set uses T-R separation distances smaller than or equal to 50 m, and the distance is larger than 50 m for the measurement set; for the InH office scenario, the prediction set corresponds to T-R separation distances smaller than or equal to 15 m, and the measurement set contains data with distances larger than 15 m, considering the generally shorter T-R separations compared to outdoor cases. The prediction results for the UMi SC scenario are illustrated in Figs.~\ref{fig:Pre_Dist_larger_UMi} and~\ref{fig:Pre_Dist_larger_para_UMi}, while Figs.~\ref{fig:Pre_Dist_larger_InH} and~\ref{fig:Pre_Dist_larger_para_InH} display the prediction performance for the InH office scenario. As shown by Figs.~\ref{fig:Pre_Dist_larger_UMi} to~\ref{fig:Pre_Dist_larger_para_InH}, the prediction error of the ABG model fluctuates significantly and rises dramatically as the measurement set gets further away from the prediction set, and may become incredibly high, e.g., over 20 dB. On the other hand, the CI and CIF models yield low (at most 8.2 dB) and very stable prediction errors across the entire range of $\delta_d$ for both UMi and InH scenarios, which implies that the CI and CIF models are both more accurate than the ABG model under varying data sets, and are not sensitive to the data set used to generate the model parameters. Similar to the UMa case, the model parameters in the CI and CIF models exhibit little variation, while the $\alpha$ and $\beta$ in the ABG model vary significantly over the investigated range of $\delta_d$. 

\begin{figure}
\centering
 \includegraphics[width=3.4in]{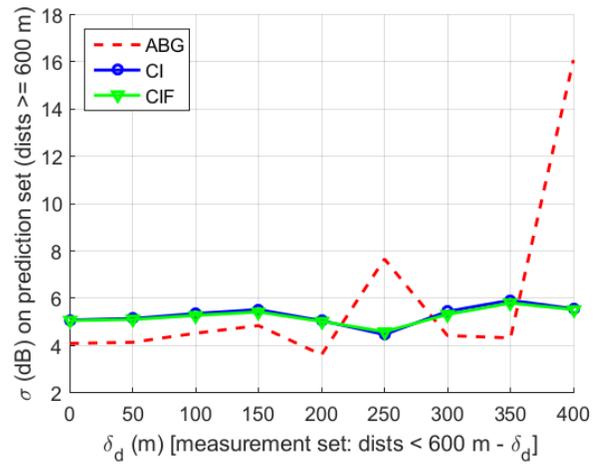}
    \caption{Shadow fading standard deviation of the ABG, CI, and CIF path loss models for prediction in distance when the measurement set is close to the transmitter in the UMa scenario.}
    \label{fig:Pre_Dist_smaller_UMa}
\end{figure}

\begin{figure}
\centering
 \includegraphics[width=3.4in]{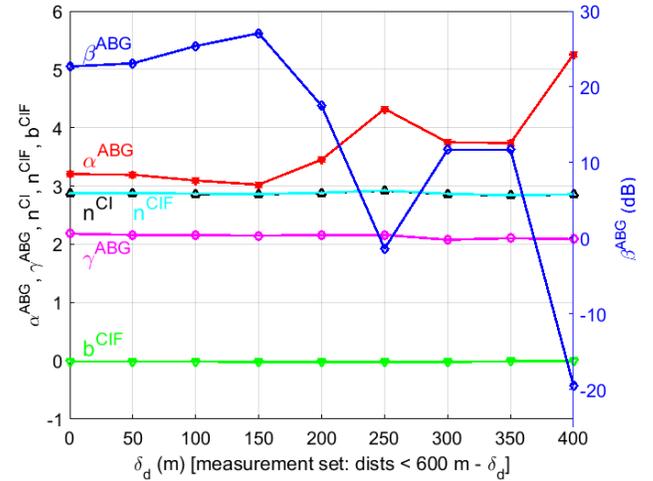}
    \caption{Parameters of the ABG, CI, and CIF path loss models for prediction in distance when the measurement set is close to the transmitter in the UMa scenario. Note that the scale for $\beta$ (dB) in the ABG model is to the right.}
    \label{fig:Pre_Dist_smaller_para_UMa}
\end{figure}

The second investigation of this experiment is for the case that the measurement set contains measured data closer to the TX (base station) than the prediction set. In this case, the prediction set contains all UMa measurements with distances larger than or equal to $d_{min}$ = 600 m, and the measurement set varies with all distances smaller than $d_{min}-\delta_d~(\delta_d \geqslant 0)$. The results for this case in the UMa scenario are shown in Fig.~\ref{fig:Pre_Dist_smaller_UMa} and Fig.~\ref{fig:Pre_Dist_smaller_para_UMa} for the SF standard deviation on the prediction set and the parameters of the path loss models, respectively, both as a function of $\delta_d$. As shown by Fig.~\ref{fig:Pre_Dist_smaller_UMa}, the prediction errors of both the CI and CIF path loss models vary very little as the distance between the measurement set and prediction set increases, while the prediction error of the ABG model on the prediction set exhibits significant variation as $\delta_d$ increases. Notice that the prediction errors of both the CI and CIF models vary by up to only 1.4 dB across the entire range of $\delta_d$ (from 0 to 400 m); in contrast, the prediction error of the ABG model can be as large as 16.1 dB and the maximum difference in prediction error reaches 12.5 dB across the entire range of $\delta_d$. Moreover, the stabilities of the modeling parameters in the CI and CIF models are much better compared to those of the ABG model when varying the distance between the two sets, as illustrated by Fig.~\ref{fig:Pre_Dist_smaller_para_UMa}, where the $\alpha$ and $\beta$ of the ABG model vary by 2.2 and 46.6 dB, respectively. This, again, shows the great sensitivity and inaccuracy (gross errors) of the ABG model to the particular data used to create the model parameters and the remarkable accuracy and robustness of the CI/CIF models to various measurement sets.

\subsection{Prediction in Frequency}
In this section, the prediction set contains the data for a given frequency and the measurement set corresponds to all the other frequencies. For example, the prediction set could be all data at 2 GHz and the measurement set the data for all the other frequencies (10, 18, 28, and 38 GHz) for the UMa scenario.

\begin{figure}
\centering
 \includegraphics[width=3.4in]{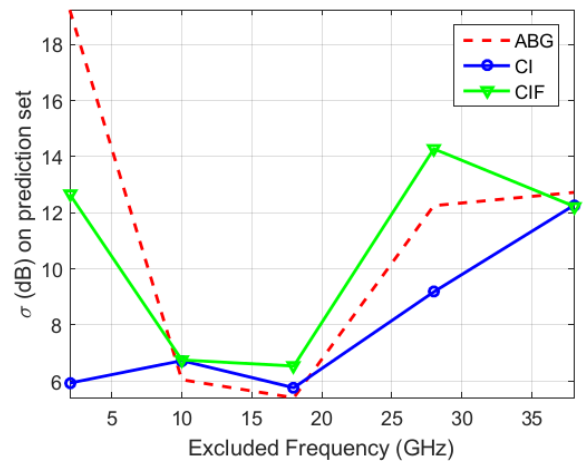}
    \caption{Shadow fading standard deviation for the ABG, CI, and CIF path loss models for prediction in frequency in the UMa scenario. The measurement set is for all frequencies except the excluded one shown on the x axis which is the prediction set.}
    \label{fig:Pre_Freq}
\end{figure}

\begin{figure}
\centering
 \includegraphics[width=3.4in]{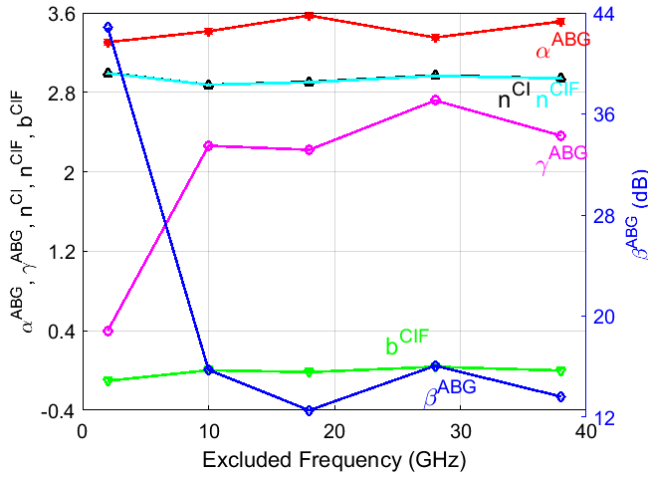}
    \caption{Parameters of the ABG, CI, and CIF path loss models for prediction in frequency in the UMa scenario. The measurement set is for all frequencies except the excluded one shown on the x axis which is the prediction set. Note that the scale for $\beta$ (dB) in the ABG model is to the right.}
    \label{fig:Pre_Freq_para}
\end{figure}

Fig.~\ref{fig:Pre_Freq} depicts the RMS error for the three path loss models on the prediction and measurement sets for the frequency shown on the x axis (where the frequency on the x axis comprises all data in the prediction set). It can be observed from Fig.~\ref{fig:Pre_Freq} that although all the three models yield varying prediction errors across the entire frequency range, the variation is the largest for the ABG model. The prediction error of the ABG model is much greater (about 19 dB) at lower frequencies where legacy 4G systems will work, showing the liability of the ABG model for simultaneous use in lower frequency and mmWave systems. The CI model shows the most robust and accurate prediction over all frequencies. The parameters of the three path loss models for prediction in frequency are shown in Fig.~\ref{fig:Pre_Freq_para}. It is obvious from Fig.~\ref{fig:Pre_Freq_para} that the parameters in the CI and CIF models vary much less across frequencies as compared to the parameters in the ABG model, demonstrating the liability of the ABG model in terms of the sensitivity analysis of specific frequencies and measurements used in the data sets. 

\subsection{Prediction Across Environments}
Fig.~\ref{fig:Pre_Freq} and Fig.~\ref{fig:Pre_Freq_para} also show the prediction performance of the three path loss models across environments, when considering an arbitrary single frequency, e.g., focusing on the results associated with 38 GHz. The 2, 10, 18, and 28 GHz data were measured in the Aalborg UMa environment, while the 38 GHz data were obtained from the Austin UMa environment, hence prediction results at 38 GHz actually show how the three path loss models behave when using the Aalborg data to predict the Austin data. As seen in Fig.~\ref{fig:Pre_Freq}, the prediction errors for the CI and CIF models at 38 GHz are slightly smaller than the ABG model, indicating that all three models yield comparable prediction performance when applied in different cities. 

These results, as well as those in\cite{Tho16:VTC}, show superior prediction ability and robust sensitivity of the CI path loss model for outdoor scenarios, and the virtue of the CIF model for indoor settings in the large majority of cases. This advantage is especially useful for 5G mmWave standardization where an accurate, trustworthy model must be developed without the benefit of a complete set of measurements across all frequencies and all environments, especially given the fact that future spectrum may be allocated in bands different from what was originally measured.

\section{Conclusion}
In this paper, we have provided a comparison of three large-scale propagation path loss models, i.e., the ABG (four parameters), CI (two parameters), and CIF (three parameters) models, over the microwave and mmWave frequency bands using 30 sets of measurement data from 2 GHz to 73 GHz for UMa, UMi, and InH scenarios. 

First, comparisons were made between the 1-m CI model and the CI model with an optimized reference distance $d_0$ (CI-opt). Results show that the two-parameter 1-m CI model provides virtually identical accuracy as compared to the three-parameter CI-opt model, and the CI-opt model can sometimes yield unrealistic PLEs. The data prove that a 1-m free-space reference distance, rather than an optimized $d_0$, is justified for the CI model.

Work here showed that the ABG, CI and CIF models are all very comparable in prediction accuracy when large data sets exist, even though the ABG model requires more model parameters and lacks a physical basis for its floating intercept value. By contrast, the CI and CIF models are physically tied to the transmitter power via the utilization of a 1-m close-in free-space reference distance that has inherent frequency dependency over the entire microwave and mmWave bands. This allows for comparable accuracy but greater parameter stability using fewer model parameters, and for easy \lq\lq{}in your head\rq\rq{} computation of mean path loss at all distances, by virtue of just a single model parameter (PLE or $n$) for the CI model (where 10$n$ is the path loss in dB per decade of distance beyond 1 m) and two model parameters ($n$ and $b$) for the CIF model. No change in mathematical form, and the change of just a single constant is all that is needed to change the existing 3GPP floating-intercept (AB/ABG) path loss model to the simpler and more stable CI/CIF models which provide virtually identical accuracy compared to the four-parameter ABG model over a vast range of frequencies --- from today's cellular to future mmWave bands. This paper showed that the AB and ABG models have parameter values that vary greatly across different frequency and distance ranges, while reducing the SF standard deviation by only a fraction of a dB in most cases compared to the physically-based CI and CIF models that use fewer model parameters. The single greatest difference between standard deviations for all three models over all scenarios was found to be 1.2 dB for the UMi scenario, where only 82 data points were available. However, a recent study with a much richer data set\cite{Han:VTC} showed only 0.4 dB difference between the ABG and CI models in UMi. 

This paper showed, by way of example at 28 GHz, that the ABG NLOS model has inherent inaccuracy at both small ($<$ 30 m) and large (several hundred meters) distances, and predicts \textit{less than free space} loss when close to the TX while underestimating interference at large distances when used at an arbitrary frequency as compared to CI. Hence, the ABG model will lead to overly optimistic capacity simulations. Especially for future small cell deployments, where dozens of neighboring BSs could produce interference, the simulation results would be vastly different between the ABG and CI/CIF models.

A key contribution of this paper was a sensitivity analysis that showed the CI and CIF models are superior to the ABG model in both stability performance and prediction accuracy (i.e., SF standard deviation) over a vast frequency range, when using the model to predict path loss at different distances and frequencies relative to the set of data from which the parameters of the path loss models were originally determined. Thus, for unexpected scenarios or for situations where a path loss model may be used at different distances or frequencies than the measurements used to create the original model, the sensitivity analysis in this paper shows the CI and CIF models are more robust, accurate, and reliable as compared to the ABG model.

Finally, the CI model was shown to be most suitable for outdoor environments because of its accuracy, simplicity, and superior sensitivity performance due to its physical close-in free space reference point, given the fact that measured path loss exhibits little dependence on frequency in outdoor environments beyond the first meter of free space propagation (captured in the FSPL term). On the other hand, the CIF model is well suited for indoor environments, since it provides a smaller standard deviation than the ABG model in many cases even with fewer model parameters, and has superior accuracy when scrutinized with the sensitivity analysis.


\section*{Appendix}
Mathematical derivations for the closed-form solutions for the ABG, CI, and CIF models, by solving for model parameters that minimize the SF standard deviation, are provided in this appendix. Note that all the frequencies are in GHz here.
\subsection{ABG Path Loss Model}
The ABG model can be expressed as (with 1 m reference distance and 1 GHz reference frequency)\cite{Pie12}:
\begin{equation}\label{ABG}
\begin{split}
\PL^{\ABG}(f,d)[\dB]=&10\alpha log_{10}(\frac{d}{1~m})+\beta+10\gamma log_{10}(\frac{f}{1~GHz})\\
&+\chi_{\sigma}^{\ABG}
\end{split}
\end{equation}

\noindent Assuming $B=\PL^{\ABG}(f,d)[\dB]$, $D=10log_{10}(d)$, and $F=10log_{10}(f)$ in~\eqref{ABG}, the SF is given by:
\begin{equation}
\chi_{\sigma}^{\ABG}=B-\alpha D-\beta-\gamma F
\end{equation}

\noindent Then the SF standard deviation is:
\begin{equation}\label{ABG_sigma}
\sigma^{\ABG} = \sqrt{\sum{{\chi_{\sigma}^{\ABG}}^2}/N}=\sqrt{\sum{(B-\alpha D-\beta-\gamma F)^2}/N}
\end{equation}

Minimizing the fitting error is equivalent to minimizing $\sum{(B-\alpha D-\beta-\gamma F)^2}$, which means its partial derivatives with respect to $\alpha$, $\beta$, and $\gamma$ should be zero, as shown by~\eqref{ABG_derA},~\eqref{ABG_derB}, and~\eqref{ABG_derG}.
\begin{equation}\label{ABG_derA}
\begin{split}
\frac{\partial \sum{(B-\alpha D-\beta-\gamma F)^2}}{\partial \alpha}=&2(\alpha\sum{D^2}+\beta\sum{D}\\
&+\gamma\sum{DF}-\sum{DB})\\
=&0
\end{split}
\end{equation}

\begin{equation}\label{ABG_derB}
\begin{split}
\frac{\partial \sum{(B-\alpha D-\beta-\gamma F)^2}}{\partial \beta}=&2(\alpha\sum{D}+N\beta+\gamma\sum{F}\\
&-\sum{B})\\
=&0
\end{split}
\end{equation}

 \begin{equation}\label{ABG_derG}
\begin{split}
\frac{\partial \sum{(B-\alpha D-\beta-\gamma F)^2}}{\partial \gamma}=&2(\alpha\sum{DF}+\beta\sum{F}\\
&+\gamma\sum{F^2}-\sum{FB})\\
=&0
\end{split}
\end{equation}

\begin{figure*}
\begin{equation}\label{A}
\alpha=\frac{(\sum{D}\sum{B}-N\sum{DB})((\sum{F})^2-N\sum{F^2})-(\sum{D}\sum{F}-N\sum{DF})(\sum{F}\sum{B}-N\sum{FB})}{((\sum{D})^2-N\sum{D^2})((\sum{F})^2-N\sum{F^2})-(\sum{D}\sum{F}-N\sum{DF})^2}
\end{equation}
\end{figure*}

\begin{figure*}
\begin{equation}\label{B}
\beta=\frac{(\sum{D}\sum{FB}-\sum{B}\sum{DF})(\sum{F}\sum{D^2}-\sum{D}\sum{DF})-(\sum{B}\sum{D^2}-\sum{D}\sum{DB})(\sum{D}\sum{F^2}-\sum{F}\sum{DF})}{((\sum{D})^2-N\sum{D^2})(\sum{D}\sum{F^2}-\sum{F}\sum{DF})+(\sum{D}\sum{F}-N\sum{DF})(\sum{F}\sum{D^2}-\sum{D}\sum{DF})}
\end{equation}
\end{figure*}

\begin{figure*}
\begin{equation}\label{G}
\gamma=\frac{(\sum{F}\sum{B}-N\sum{FB})((\sum{D})^2-N\sum{D^2})-(\sum{D}\sum{F}-N\sum{DF})(\sum{D}\sum{B}-N\sum{DB})}{((\sum{F})^2-N\sum{F^2})((\sum{D})^2-N\sum{D^2})-(\sum{D}\sum{F}-N\sum{DF})^2}
\end{equation}
\end{figure*}

\noindent It is found from~\eqref{ABG_derA},~\eqref{ABG_derB}, and~\eqref{ABG_derG} that
\begin{equation}\label{eqA}
\alpha\sum{D^2}+\beta\sum{D}+\gamma\sum{DF}-\sum{DB}=0
\end{equation}

\begin{equation}\label{eqB}
\alpha\sum{D}+N\beta+\gamma\sum{F}-\sum{B}=0
\end{equation}

\begin{equation}\label{eqG}
\alpha\sum{DF}+\beta\sum{F}+\gamma\sum{F^2}-\sum{FB}=0
\end{equation}

\noindent Through calculation and simplification, we obtain the closed-form solutions for $\alpha$, $\beta$, and $\gamma$ as shown by~\eqref{A},~\eqref{B}, and~\eqref{G}, respectively. Finally, the minimum SF standard deviation for the ABG model can be obtained by plugging~\eqref{A},~\eqref{B}, and~\eqref{G} back into~\eqref{ABG_sigma}.

\subsection{CI Path Loss Model with Optimized Free Space Reference Distance}\label{subsec:CI_opt}
The expression for the CI model with a reference distance of $d_0$ is given by\cite{Rap15:TCOM}:
\begin{equation}\label{CI2}
\begin{split}
\PL^{\CI}(f,d)[\dB]=&20\log_{10}\left(\frac{4\pi fd_0\times 10^9}{c}\right)+10n\log_{10}\left(\frac{d}{d_0}\right)\\
&+\chi_{\sigma}^{\CI}\\
=&20\log_{10}\left(\frac{4\pi f\times 10^9}{c}\right)+20\log_{10}(d_0)\\
&+10n\log_{10}(d)-10n\log_{10}(d_0)+\chi_{\sigma}^{\CI}\\
\end{split}
\end{equation}

\noindent Thus the SF is:
\begin{equation}
\begin{split}
\chi_{\sigma}^{\CI}=&\PL^{\CI}(f,d)[\dB]-20\log_{10}\left(\frac{4\pi f\times 10^9}{c}\right)-20\log_{10}(d_0)\\
&-10n\log_{10}(d)+10n\log_{10}(d_0)\\
\end{split}
\end{equation}

Let $A=\PL^{\CI}(f,d)[\dB]-20\log_{10}\left(\frac{4\pi f\times 10^9}{c}\right)$, $B=10\log_{10}(d_0)$, $D=10\log_{10}(d)$, then we have
\begin{equation}
\begin{split}
\chi_{\sigma}^{\CI}=&A-2B-nD+nB=A-nD-(2-n)B\\
=&A-nD-b
\end{split}
\end{equation}

\noindent where $b=(2-n)B$. Then the SF standard deviation is:
\begin{equation}
\sigma^{\CI} = \sqrt{\sum{{\chi_{\sigma}^{\CI}}^2}/N}=\sqrt{\sum{(A-nD-b)^2}/N}
\end{equation}

\noindent where $N$ is the number of path loss data points. Thus minimizing the SF standard deviation $\sigma^{CI}$ is equivalent to minimizing the term $\sum{(A-nD-b)^2}$. When $\sum{(A-nD-b)^2}$ is minimized, its derivatives with respect to $n$ and $b$ should be zero, i.e.,
\begin{equation}\label{n_der}
\begin{split}
\frac{d\sum{(A-nD-b)^2}}{dn}&=\sum{2D(nD+b-A)}=0\\
\end{split}
\end{equation}

\begin{equation}\label{b_der}
\begin{split}
\frac{d\sum{(A-nD-b)^2}}{db}&=\sum{2(nD+b-A)}=0\\
\end{split}
\end{equation}

\noindent By jointly solving~\eqref{n_der} and~\eqref{b_der} we can obtain
\begin{equation}\label{n}
n=\frac{\sum A\sum D-N\sum{DA}}{(\sum D)^2-N\sum{D^2}}
\end{equation}

\begin{equation}\label{b}
b=\frac{\sum A-n\sum D}{N}
\end{equation}

\noindent i.e.,
\begin{equation}\label{d_0}
d_0=10^{B/10}=10^{\frac{\sum A-n\sum D}{10N(2-n)}}
\end{equation}

\subsection{CI Path Loss Model with 1 m Free Space Reference Distance}\label{subsec:CI_1m}
The expression for the CI model with a reference distance of 1 m is given by\cite{Rap15:TCOM}:
\begin{equation}\label{CI}
\PL^{\CI}(f,d)[\dB]=\FSPL(f, 1~m)[\dB]+10nlog_{10}(d)+\chi_{\sigma}^{\CI}
\end{equation}

\noindent where
\begin{equation}\label{FSPL1}
\FSPL(f, 1~m)[\dB]=20\log_{10}\left(\frac{4\pi f\times 10^9}{c}\right)
\end{equation}

\noindent Thus the SF is:
\begin{equation}
\begin{split}
\chi_{\sigma}^{\CI}&=\PL^{\CI}(f,d)[\dB]-\FSPL(f, 1~m)[\dB]-10nlog_{10}(d)\\
&=A-nD
\end{split}
\end{equation}

\noindent where $A$ represents $\PL^{\CI}(f,d)[dB]-\FSPL(f, 1~m)[dB]$, and $D$ denotes $10log_{10}(d)$. Then the SF standard deviation is:
\begin{equation}
\sigma^{\CI} = \sqrt{\sum{{\chi_{\sigma}^{\CI}}^2}/N}=\sqrt{\sum{(A-nD)^2}/N}
\end{equation}

\noindent where $N$ is the number of path loss data points. Thus minimizing the SF standard deviation $\sigma^{CI}$ is equivalent to minimizing the term $\sum{(A-nD)^2}$. When $\sum{(A-nD)^2}$ is minimized, its derivative with respect to $n$ should be zero, i.e.,
\begin{equation}\label{PLE_der}
\begin{split}
\frac{d\sum{(A-nD)^2}}{dn}&=\sum{2D(nD-A)}=0\\
\end{split}
\end{equation}

\noindent Therefore, from~\eqref{PLE_der} we have
\begin{equation}\label{PLE}
n=\frac{\sum{DA}}{\sum{D^2}}
\end{equation}

\subsection{CIF Path Loss Model}
The equation of the CIF model~\eqref{CIF1} with a reference distance of 1 m is re-organized in the form:
\begin{equation}\label{CIF}
\begin{split}
\PL^{\CIF}(f,d)[\dB]=
&\FSPL(f, 1~\textrm{m})[\dB]\\
&+10\log_{10}(d)(n(1-b)+\frac{nb}{f_0}f)+X_{\sigma}^{\CIF}
\end{split}
\end{equation}

\noindent where $n$ is the PLE that includes the frequency-effect parameter $b$, and $f_0$ is the specified reference frequency that may be selected as the average of all measured frequencies. Let $A=\PL^{\CIF}(f,d)[\dB]-\FSPL(f, 1~\textrm{m})[\dB]$, $D=10\log_{10}(d)$, $a=n(1-b)$, and $g=\frac{nb}{f_0}$, then we have:
\begin{equation}
X_{\sigma}^{\CIF}=A-D(a+gf)
\end{equation}
The SF standard deviation is:
\begin{equation}\label{CIF_sigma}
\sigma^{\CIF} = \sqrt{\sum{{X_{\sigma}^{\CIF}}^2}/N}=\sqrt{\sum{(A-D(a+gf))^2}/N}
\end{equation}
Minimizing $\sigma^{\CIF}$ is equivalent to minimizing $\sum{(A-D(a+gf))^2}$. When $\sum{(A-D(a+gf))^2}$ is minimized, its derivatives with respect to $a$ and $g$ should be zero, i.e.
\begin{equation}
\begin{split}
\frac{\partial \sum{(A-D(a+gf))^2}}{\partial a}=&\sum{2D(aD+gDf-A)}\\
=&2(a\sum{D^2}+g\sum{D^2f}-\sum{DA})\\
=&0
\end{split}
\end{equation}

\begin{equation}
\begin{split}
\frac{\partial \sum{(A-D(a+gf))^2}}{\partial g}=&\sum{2Df(aD+gDf-A)}\\
=&2(a\sum{D^2f}+g\sum{D^2f^2}\\
&-\sum{DAf})\\
=&0
\end{split}
\end{equation}
which can be simplified to:
\begin{equation}\label{CIF_aDer}
a\sum{D^2}+g\sum{D^2f}-\sum{DA}=0
\end{equation}

\begin{equation}\label{CIF_gDer}
a\sum{D^2f}+g\sum{D^2f^2}-\sum{DAf}=0
\end{equation}
Combining~\eqref{CIF_aDer} and~\eqref{CIF_gDer} yields:
\begin{equation}\label{CIF_a}
a=\frac{\sum{D^2f}\sum{DAf}-\sum{D^2f^2}\sum{DA}}{(\sum{D^2f})^2-\sum{D^2}\sum{D^2f^2}}
\end{equation}

\begin{equation}\label{CIF_g}
g=\frac{\sum{D^2f}\sum{DA}-\sum{D^2}\sum{DAf}}{(\sum{D^2f})^2-\sum{D^2}\sum{D^2f^2}}
\end{equation}

\noindent Put into matrix form, $a$ and $g$ are:
\begin{equation}\label{CIF_aM}
\small a=\frac{f^T\diag(DD^T)f^T\diag(DA^T)-(\diag(ff^T))^T\diag(DD^T)D^TA}{(f^T\diag(DD^T))^2-(\diag(ff^T))^T\diag(DD^T)D^TD}
\end{equation}

\begin{equation}\label{CIF_gM}
\small g=\frac{f^T\diag(DD^T)D^TA-f^T\diag(DA^T)D^TD}{(f^T\diag(DD^T))^2-(\diag(ff^T))^T\diag(DD^T)D^TD}
\end{equation}
Equations~\eqref{CIF_a}--\eqref{CIF_gM} are closed-form solutions for $a$ and $g$. Substituting $a$ and $g$ in~\eqref{CIF_sigma} with~\eqref{CIF_aM} and~\eqref{CIF_gM}, the minimum SF standard deviation for the CIF model is found. 

After solving for $a$ and $g$, we can use the previous definition $a=n(1-b)$ and $g=\frac{nb}{f_0}$ to calculate $n$, $b$, and $f_0$. However, there are two equations but three unknowns, hence there is no unique solution in general using three parameters. However, a unique closed-form solution is available when $f_0$ is specified as a constant deemed appropriate by the user, such as the weighted average of all frequencies used in the model, or at a natural loss transition band (e.g., where measurements show an inflection point in the PLE), or at known transition points like the 60 GHz oxygen absorption band. Consequently, $n$ and $b$ are solved by:
\begin{equation}
n=a+gf_0
\end{equation}

\begin{equation}
b=\frac{gf_0}{a+gf_0}
\end{equation}

\ifCLASSOPTIONcaptionsoff
  \newpage
\fi

\bibliographystyle{IEEEtran}
\bibliography{bibliography}

\end{document}